\let\accentvec\vec

\documentclass[graybox]{svmult}

\usepackage{mathptmx}       
\usepackage{helvet}         
\usepackage{courier}        
\usepackage{type1cm}        
\usepackage{makeidx}         
\usepackage{graphicx}        
\usepackage{multicol}        
\usepackage[bottom]{footmisc}
\usepackage{amsmath,amssymb,bbm}
\usepackage{cite}

\makeindex             
                       
\newcommand{\be}{\begin{equation}}
\newcommand{\ee}{\end{equation}}
 
\newcommand{\ba}{\begin{eqnarray}}
\newcommand{\ea}{\end{eqnarray}}

\newcommand{\bas}{\begin{eqnarray*}}
\newcommand{\eas}{\end{eqnarray*}}

\newcommand{\bc}{\begin{center}}
\newcommand{\ec}{\end{center}}

\newcommand{\bay}{\begin{array}{rcl}}
\newcommand{\eay}{\end{array}}
 
\newcounter{subequation}[equation]
\makeatletter

\expandafter\let\expandafter
\reset@font\csname reset@font\endcsname

\def\subeqnarray{\arraycolsep1pt
    \def\@eqnnum\stepcounter##1{\stepcounter{subequation}%
        {\reset@font\rm(\theequation\alph{subequation})}}
\jot5mm     \eqnarray}

\def\subarray{\arraycolsep1pt
    \def\@eqnnum\stepcounter##1{\stepcounter{subequation}%
        {\reset@font\rm(\alph{subequation})}}
\jot5mm     \eqnarray}

\makeatother

\newcommand{\Zom}{\mathbb{Z}}

\newcommand{\ra}{\rightarrow}

\newcommand{\is}{ &\! =\! & }

\newcommand{\half}{{\textstyle{\frac{1}{2}}}}

\newcommand{\Tr}{{\rm Tr}}

\newcommand{\ie}{{\it i.e.\ }}

\newcommand{\lb}{\lambda}

\newcommand{\dd}{{\partial}}

\newcommand{\lbar}{\bar{\lambda}}

\newcommand{\cB}{{\cal B}}

\newcommand{\cD}{{\cal D}}
\newcommand{\cE}{{\cal E}}

\newcommand{\cO}{{\cal O}}

\newcommand{\cR}{{\cal R}}
\newcommand{\cS}{{\cal S}}

\newcommand{\p}{\partial}

\newcommand{\gb}{\overline{\rm g}}

\newcommand{\ub}{\bar{u}}

\newcommand{\mn}{{\mu\nu}}

\newcommand{\m}{\mu}
\newcommand{\e}{\mathrm{e}}

\def\ie{{i.e. }}
\def\lp{\ell_{\rm Pl}}
\def\LH{\ell_{\rm H}}

\def\tr{t_{\rm tr}}
\def\mp{m_{\rm Pl}}
\def\tp{t_{\rm Pl}}
\def\OM{\Omega_{\rm M}}
\def\OL{\Omega_{\bar{\lambda}}}

\def\OLS{\Omega_{\bar{\lambda}}^\ast}

\def\barg{\bar{G}}
\def\lat{\lambda_{\rm T}}

\def\kat{k_{\rm T}}
\def\gat{g_{\rm T}}
\def\hT{H_{\rm T}}
\def\h0{H_{0}}
\def\aT{a_{\rm T}}

\def\tT{t_{\rm T}}
 
\def\calp{\widetilde{\cal P}}
\def\clp{{\cal P}}

\def\UU{{\cal U}}
\def\ssc{S_{\rm c}}

\begin{document}

\thispagestyle{empty}
\begin{flushright} \small
MZ-TH/12-02
\end{flushright}
\bigskip

\begin{center}
 {\LARGE\bfseries   Asymptotic Safety, Fractals, \\ and Cosmology\footnote{Lectures given by M.R.\ at the Sixth Aegean Summer School on Quantum Gravity
and Quantum Cosmology, Chora, Naxos (Greece), September 2011.}\\[15mm] 
}

Martin Reuter and Frank Saueressig \\[3mm]
{\small\slshape
Institute of Physics, University of Mainz\\
Staudingerweg 7, D-55099 Mainz, Germany \\[1.1ex]
{\upshape\ttfamily reuter@thep.physik.uni-mainz.de} \\
{\upshape\ttfamily saueressig@thep.physik.uni-mainz.de} }\\
\end{center}
\vspace{10mm}

\hrule\bigskip

\centerline{\bfseries Abstract} \medskip
\noindent
These lecture notes introduce the basic ideas of the Asymptotic Safety approach to Quantum
	Einstein Gravity (QEG). In particular they provide the background for recent work on the possibly
	multifractal structure of the QEG space-times. Implications of Asymptotic Safety for the cosmology
	of the early Universe are also discussed.
\bigskip
\hrule\bigskip
\newpage







\section{Introduction}
\label{reu:sect:1}
Finding a consistent and fundamental quantum theory
for gravity is still one of the most challenging
open problems in theoretical high energy physics to date.
As is well known, the perturbative quantization of the classical description for gravity, 
General Relativity, results in a non-renormalizable quantum theory \cite{reu:tHooft:1974bx,reu:Goroff:1985sz,reu:vandeVen:1991gw}. 
One possible lesson drawn from this result may assert that gravity constitutes an effective field theory valid at low energies, whose UV completion requires 
the introduction of new degrees of freedom and symmetries. This is the path followed, e.g., by string
theory. In a less radical approach, one retains the fields and symmetries known from
General Relativity and conjectures that gravity constitutes a fundamental theory
at the non-perturbative level. One proposal along this line is
the Asymptotic Safety scenario \cite{reu:livrev}
initially put forward by Weinberg \cite{reu:wein,reu:Weinproc1,reu:Weinproc2}.
The key ingredient in this scenario 
is a non-Gaussian fixed point (NGFP) of the gravitational renormalization group (RG) flow, 
which controls the behavior of the theory at high energies and renders physical quantities safe from unphysical divergences. 
Given that the NGFP comes with a finite number of unstable (or relevant) directions 
this construction is as predictive as a ``standard'' perturbatively renormalizable quantum field theory.

\noindent
{\bf (1)} The primary tool for investigating this scenario is the functional renormalization 
group equation (FRGE) for gravity \cite{reu:mr}, which constitutes the spring-board
for the detailed 
investigations of the non-perturbative renormalization 
group (RG)\index{RG}\index{renormalization group|see{RG}} behavior of Quantum Einstein Gravity \cite{reu:mr,reu:percadou,reu:oliver1,reu:frank1,reu:oliver2,reu:oliver3,reu:oliver4,reu:souma,reu:frank2,reu:prop,reu:perper1,reu:codello,reu:litimgrav,reu:essential,reu:r6,reu:hier,reu:MS1,reu:Codello:2008vh,reu:creh1,reu:creh2,reu:creh3,reu:JE1,reu:HD1,reu:HD2,reu:JEUM,reu:Eichhorn:2009ah,reu:Groh:2010ta,reu:Eichhorn:2010tb,reu:elisa2,reu:MRS,reu:elisa1,reu:frank+friends,reu:Benedetti:2010nr,reu:e-omega,reu:maxpert,reu:Saueressig:2011vn,reu:Benedetti:2011ct,reu:max}.
The FRGE defines a Wilsonian RG flow on a theory space
which consists of all diffeomorphism invariant 
functionals of the metric $g_{\mu \nu}$, 
and turned out to be ideal for 
investigating the Asymptotic Safety conjecture
 \cite{reu:wein,reu:livrev}.
In fact, it yielded substantial evidence for 
the non-perturbative renormalizability
of Quantum Einstein Gravity\index{Quantum Einstein Gravity|see{QEG}}. The theory emerging 
from this construction (henceforth denoted
``QEG''\index{QEG}) is not a quantization of classical 
General Relativity. Instead, its bare action corresponds to
a non-trivial fixed point of the RG flow and is a prediction therefore. 

The approach of \cite{reu:mr} employs the effective average
action\index{effective average action} $\Gamma_k$ \cite{reu:avact,reu:ym,reu:avactrev,reu:ymrev}
which has crucial advantages as compared to other
continuum implementations of the Wilsonian RG flow \cite{reu:bagber}. 
The scale dependence of $\Gamma_k$ is governed by the FRGE \cite{reu:avact}
\be\label{reu:E9}
k \p_k \Gamma_k[\Phi, \bar{\Phi}] = \frac{1}{2} {\rm STr} \left[ 
\left( \frac{\delta^2 \Gamma_k}{\delta \Phi^A \delta \Phi^B} + \cR_k \right)^{-1} k \p_k \cR_k 
\right] \, . 
\ee
Here $\Phi^A$ is the collection of all dynamical fields considered, $\bar{\Phi}^A$ denotes their background
counterparts\index{background field} and STr denotes a generalized
functional trace carrying a minus sign for fermionic fields and a factor 2 for complex fields.
Moreover $\cR_k$ is a matrix-valued infrared cutoff\index{cutoff (infrared)}, which provides a $k$-dependent mass-term for
fluctuations with momenta $p^2 \ll k^2$, while vanishing for $p^2 \gg k^2$.
Solutions of the flow equation give rise to 
 families of effective field theories $\{ \Gamma_k[g_{\mu \nu}], 0 \le k < \infty \}$ 
labeled by the coarse graining scale $k$. 
The latter property opens the door to a rather direct extraction of physical 
information from the RG flow, at least in single-scale cases: If the physical
 process under consideration involves 
 a single typical momentum scale $p_0$ only, it can be described by a tree-level 
evaluation of $\Gamma_k[g_{\mu \nu}]$, with $k = p_0$.

\noindent
{\bf (2)} Already soon after the Asymptotic Safety program had taken its modern form,
various indications pointed in the direction that in QEG space-time should have certain features in common with a fractal. In ref.\ \cite{reu:oliver1} the four-dimensional graviton propagator has been studied in the regime of asymptotically large momenta and it has been found that near the Planck scale\index{Planck scale} a kind of dynamical dimensional reduction occurs. As a consequence of the NGFP\index{NGFP} controlling the UV behavior\index{UV behavior} of the theory, the
four-dimensional graviton propagator essentially behaves two-dimensional on microscopic scales.

Subsequently, the "finger prints" of the NGFP on the fabric of the effective QEG space-times have been discussed in \cite{reu:oliver2}, where it was shown that Asymptotic Safety induces a characteristic self-similarity of space-time on length-scales below the Planck length $\ell_{\rm PL}$. The graviton propagator becomes scale-invariant in this regime \cite{reu:oliver1}. Based on this observation it was argued that, in a cosmological context, the geometry fluctuations it describes can give rise to a scale free spectrum of primordial density perturbations responsible for structure formation \cite{reu:cosmo1,reu:entropy}. Thus the overall picture of the space-time structure in asymptotically safe gravity as it emerged about ten years ago comprises a smooth classical manifold on large distance scales, while on small scales one encounters a low dimensional effective fractal \cite{reu:oliver1,reu:oliver2}.

The characteristic feature at the heart of these results is that the effective field equations derived from the gravitational average action equip every given smooth space-time manifold with, in principle, infinitely  many different (pseudo) Riemannian structures, one for each coarse graining scale \cite{reu:jan1,reu:jan2}. Thus, very much like in the famous example of the coast line of England \cite{reu:mandel}, the proper length on a QEG space-time depends on the "length of the yardstick" used to measure it. Earlier on similar fractal properties had already been found in other quantum gravity theories, in particular near dimension 2 \cite{reu:ninomiya}, in a non-asymptotically safe model \cite{reu:percacci-floreanini-frac} and by analyzing the conformal anomaly \cite{reu:nino}.

Along a different line of investigations, the Causal Dynamical Triangulation (CDT) approach has been developed and first Monte-Carlo simulations were performed \cite{reu:ajl1,reu:ajl2,reu:ajl34,ajl5,reu:Benedetti:2009ge,reu:Kommu:2011wd}, see \cite{reu:Ambjorn:2009ts} for a recent review. In this framework one attempts to compute quantum gravity partition functions by numerically constructing the continuum limit of an appropriate statistical mechanics system. This limit amounts to a second order phase transition. If CDT and its counterpart QEG, formulated in the continuum by means of the average action, belong to the same universality\index{universality} class one may expect that the phase transition of the former is described by the non-trivial fixed point underlying the Asymptotic Safety of the latter.

Remarkably, ref. \cite{reu:ajl34} reported results which indicated that the four-dimensional CDT space-times, too, undergo a dimensional reduction from four to two dimensions as one ``zooms'' in on short distances. In particular it had been demonstrated that the spectral dimension $d_s$ measured in the CDT simulations has the very same limiting behaviors, $4 \rightarrow 2$, as in QEG \cite{reu:oliverfrac}. Therefore it was plausible to assume that both approaches indeed ``see'' the same continuum physics. 

This interpretation became problematic when  
ref.\ \cite{reu:Benedetti:2009ge} carried out CDT simulations for $d=3$ macroscopic dimensions, which favor a value near $d_s=2$ on the shortest length-scale probed since, in this case, the QEG prediction for the fixed point region is the value $d_s=3/2$ \cite{reu:oliverfrac}. Furthermore, the authors of ref.\ \cite{reu:laiho-coumbe} reported simulations within the {\it euclidean} dynamical triangulation (EDT) approach in $d=4$, which favor a drop of the spectral dimension from 4 to about 1.5; this is again in conflict with the QEG expectations if one interprets the latter dimension as the value in the continuum limit.

Later on we will present several types of scale dependent effective dimensions, specifically the spectral dimension $d_s$ and the walk dimension $d_w$ for the effective QEG space-times. We shall see that on length scales slightly {\it larger} than $\ell_{\rm PL}$ there exists a further regime which exhibits the phenomenon of dynamical dimensional reduction. There the spectral dimension is even smaller than near the fixed point, namely $d_s=4/3$ in the case of 4 dimensions classically. Moreover, we shall argue that the (3-dimensional) results reported in \cite{reu:Benedetti:2009ge} are in perfect accord with QEG, but that the shortest possible length scale achieved in the simulations is not yet close to the Planck length. Rather the Monte Carlo data probes the transition between the classical and the newly discovered ``semi-classical'' regime \cite{reu:frankfrac}.

For similar work on fractal features in different approaches we must refer to the literature
\cite{reu:modesto2008,reu:Modesto:2009kq,reu:modesto-caravelli2009,reu:carlip,reu:ncgeom,reu:spec-trip-frac,reu:Dario-kappa,reu:Horavafit,
reu:calcagni-applications,reu:calcagni-etal,reu:calcagni-reviews,reu:dunne-phot,reu:dunne-complexdim,reu:hill}.

\noindent
{\bf (3)} As for possible physics implications of the RG flow 
predicted by QEG, ideas from particle physics, in 
particular the ``RG improvement'', have been employed 
in order to study the leading quantum gravity effects 
in black holes \cite{reu:bh,reu:erick1}, cosmological space-times
\cite{reu:cosmo1,reu:cosmo2,reu:elo,reu:entropy,reu:wein3,reu:h1,reu:h2,reu:h3,reu:Ward:2008sm,reu:Bonanno:2010bt,Hindmarsh:2011hx}
or possible observable signatures from Asymptotic Safety at the LHC \cite{reu:Hewett:2007st,reu:Litim:2007iu,reu:Koch:2007yt,reu:Falls:2010he}.
Among other results, it was found \cite{reu:bh} that the quantum 
effects tend to decrease the Hawking temperature of black holes, 
and that their evaporation process presumably stops 
completely once the black holes mass is of the order of the Planck mass.
In cosmology it turned out that inflation can occur without the need of an
inflaton, and that the running of the cosmological constant might be responsible
for the observed entropy of the present Universe \cite{reu:entropy}.\smallskip

These lectures are intended to provide the necessary background for these developments.
They consist of three main parts, dealing with the basic ideas of Asymptotic Safety, the
fractal QEG space-times, and possible implications of Asymptotic Safety for cosmology,
respectively.

\section{Theory space and its truncation}
\label{reu:sect:2}

\begin{figure}[t]
\sidecaption
\includegraphics[width=.64\columnwidth]{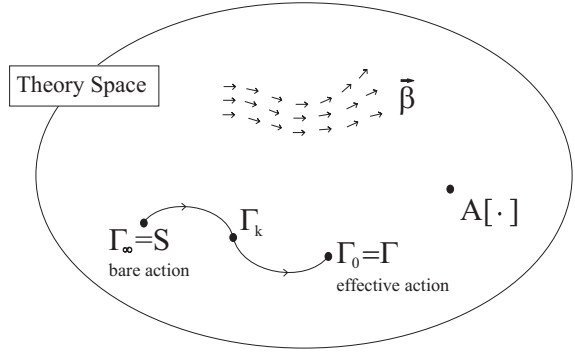}
\caption{The points of theory space are the action functionals $A[\, \cdot \,]$. The RG equation defines a vector field
$\accentvec{\beta}$ on this space; its integral curves are the RG trajectories $k \mapsto \Gamma_k$. They emanate from the fixed
point action $\Gamma_* \equiv \Gamma_\infty$, which might differ from the bare action by a simple explicitly known functional, and end at the standard effective action $\Gamma$.}
\label{reu:theoryspace}
\end{figure}

\index{truncation|textbf}
We start by reviewing the basic ideas underlying Asymptotic Safety, referring 
to \cite{reu:livrev} for a more detailed discussion.
The arena in which the Wilsonian RG\index{RG} dynamics takes place is ``theory space''\index{theory space|textbf}.
Albeit a somewhat formal notion it helps in visualizing various concepts related to functional renormalization
group equations\index{FRGE}\index{functional renormalization group equation|see{FRGE}},
see fig.\ \ref{reu:theoryspace}. To describe it, we shall consider
an arbitrary set of fields $\Phi(x)$. Then the corresponding theory space
consists of all (action) functionals $A: \Phi \mapsto A[\Phi]$ depending on this
set, possibly subject to certain symmetry requirements (a $\Zom_2$-symmetry
for a single scalar, or diffeomorphism  invariance if $\Phi$ denotes the space-time
metric, for instance). So the theory space $\{A[\, \cdot \,]\}$ is completely determined once the field 
content and the symmetries are fixed. Let us assume we can find a set of ``basis functionals''
$\{ P_\alpha[ \, \cdot \, ] \}$ so that every point of theory space has an expansion of the form 
\be\label{reu:Aexpansion}
A[\Phi, \bar{\Phi}] = \sum_{\alpha = 1}^\infty \, \ub_\alpha \, P_\alpha [\Phi, \bar{\Phi}] \, .
\ee
The basis $\{ P_\alpha[ \, \cdot \, ] \}$ will include both local field monomials and non-local
invariants and we may use the ``generalized couplings'' $\{ \bar u_\alpha , \alpha = 1,2, \cdots \}$
as local coordinates. More precisely, the theory space is coordinatized by the subset of 
``essential couplings'', i.e., those coordinates which cannot be absorbed by a field reparameterization.

Geometrically speaking the FRGE for the effective average action\index{effective average action}, eq.\ \eqref{reu:E9},
defines a vector field $\accentvec{\beta}$ on theory space. The integral curves along this vector field are the ``RG trajectories''\index{RG trajectory|textbf}
$k \mapsto \Gamma_k$ parameterized by the scale $k$. They start, for $k \ra \infty$, at the microscopic action $S$
\index{bare action} and
terminate at the ordinary effective action\index{effective action} at $k=0$. The natural orientation of the trajectories is
from higher to lower scales $k$, the direction of increasing ``coarse graining''. Expanding $\Gamma_k$ as in
\eqref{reu:Aexpansion},
\be\label{reu:Gexpansion}
\Gamma_k[\Phi, \bar{\Phi}] = \sum_{\alpha = 1}^\infty \, \ub_\alpha(k) \, P_\alpha [\Phi, \bar{\Phi}] \, ,
\ee
the trajectory is described by infinitely many ``running couplings'' $\ub_\alpha(k)$. Inserting \eqref{reu:Gexpansion} into the FRGE we obtain a system of infinitely many coupled differential equations for the $\ub_\alpha$'s:
\be\label{reu:rgeqn1}
k \partial_k \, \ub_\alpha(k) = \overline{\beta}_\alpha(\ub_1 , \ub_2 , \cdots ; k) \; , \quad \alpha = 1,2,\cdots \, .
\ee
Here the ``beta functions'' $\overline{\beta}_\alpha$ arise by 
expanding the trace on the RHS of the FRGE in terms of $\{ P_\alpha[\, \cdot \, ] \}$, i.e.,
$\tfrac{1}{2} \Tr \left[ \cdots \right] = \sum_{\alpha = 1}^\infty \overline{\beta}_\alpha(\ub_1 , \ub_2 , \cdots ; k) P_\alpha[\Phi, \bar{\Phi}]$. The expansion coefficients $\overline{\beta}_\alpha$ have the 
interpretation of beta functions similar to those of perturbation 
theory, but not restricted to relevant couplings. In standard field theory
jargon one would refer to $\ub_\alpha(k =\infty)$ as the ``bare'' parameters and to 
$\ub_\alpha(k =0)$ as the ``renormalized'' or ``dressed'' parameters. 
   
The notation with the bar on $\ub_\alpha$ and $\overline{\beta}_\alpha$ 
is to indicate that we are still dealing with dimensionful 
couplings. Usually the flow equation\index{flow equation} is reexpressed in terms of the 
dimensionless couplings 
\be\label{reu:dimless}
u_\alpha \equiv k^{-d_\alpha} \ub_\alpha, 
\ee
where $d_\alpha$ is the canonical mass dimension of $\ub_\alpha$. Correspondingly the essential $u_\alpha$'s are used as 
coordinates of theory space. The resulting 
RG\index{RG} equations 
\be 
k \dd_k u_\alpha(k) = \beta_\alpha(u_1, u_2, \cdots ) 
\label{reu:E17}
\ee
are a coupled system of autonomous differential equations. 
The $\beta_\alpha$'s have no explicit $k$-dependence and define 
a ``time independent'' vector field on theory space. 

In this language, the basic idea of renormalization can be 
understood as follows. The boundary of theory space depicted in Fig.\ \ref{reu:theoryspace} is meant to separate points with coordinates $\{u_\alpha, \alpha = 1,2,\cdots\}$ with all the essential couplings $u_\alpha$ well defined, from points with undefined, divergent couplings. The basic task of renormalization theory consists in constructing an ``infinitely long'' RG trajectory
\index{RG trajectory} which lies entirely within this theory space, i.e., a trajectory which neither leaves theory space (that is, develops divergences) in the UV limit $k \rightarrow \infty$ nor in the IR limit $k \rightarrow 0$. Every such trajectory defines one possible quantum theory.

The consistent UV-behavior can be ensured by performing the limit $k \rightarrow \infty$ at a fixed
point\index{fixed point|)}\index{fixed point|seealso{NGFP}}
$\{u_\alpha^*, \alpha = 1,2,\cdots\} \equiv u^*$ of the RG flow. The fixed point is a zero of the vector field
$\accentvec{\beta} \equiv (\beta_\alpha)$, i.e., $\beta_\alpha(u^*) = 0$ for all $\alpha = 1,2,\cdots$. The RG trajectories,
solutions of $k \partial_k u_\alpha(k) = \beta_\alpha(u(k))$, have a low ``velocity'' near a fixed point because the $\beta_\alpha$'s are small there and directly at the fixed point the running stops completely. As a result, one can ``use up'' an infinite amount of  RG time near/at the fixed point if one bases the quantum theory on a trajectory which runs into such a fixed point for $k \rightarrow \infty$. This construction ensures that in the UV limit the trajectory ends at an ``inner point'' of theory space giving rise to a well behaved action functional. Thus we can be sure that, for $k \rightarrow \infty$, the trajectory does not does not develop pathological properties such as divergent couplings. The resulting quantum theory is ``safe'' from unphysical divergences.

At this stage it is natural to distinguish two classes of fixed points. First, the UV limit may be
performed at a Gaussian fixed point (GFP)\index{Gaussian fixed point (GFP)} where $u^*_\alpha = 0, \forall \alpha = 1,2,\cdots$. 
In this case, the fixed point functional does not contain interactions and the theory becomes
asymptotically free in the UV. This is the construction underlying perturbatively renormalizable
quantum field theories. More general, one can also use a non-Gaussian fixed point (NGFP)\index{NGFP|textbf}\index{non-Gaussian fixed point|see{NGFP}} for letting
$k \rightarrow \infty$, where, by definition, not all of the coordinates $u^*_\alpha$ vanish.
In the context of gravity, Weinberg \cite{reu:wein}
proposed that the UV limit of the theory is provided by such a (NGFP). 

Note that at the NGFP it is the {\it dimensionless} essential couplings \eqref{reu:dimless}, which assume constant values. Therefore, even directly at a NGFP where $u_\alpha(k) \equiv u^*_\alpha$, the dimensionful couplings keep running according to a power law
\be
\ub_\alpha(k) = u_\alpha^* \, k^{d_{\alpha}} \,.
\ee
Furthermore, non-essential dimensionless couplings are not required to attain fixed point values.

Given a fixed point, an important concept is its {\it UV critical hypersurface} $\cS_{\rm UV}$, or synonymously, its {\it unstable manifold}. By definition, it consists of all points of theory space which are pulled into the fixed point by the inverse RG flow, i.e., for {\it in}creasing $k$.
Its dimensionality ${\rm dim}\left({\cal S}_{\rm UV}\right)\equiv \Delta_{\rm UV}$
is given by the number of attractive (for {\it in}creasing cutoff $k$) 
directions in the space of couplings.

For the RG equations \eqref{reu:E17}, 
the linearized flow near the fixed point is governed by the Jacobi matrix
${\bf B}=(B_{\alpha \gamma})$, $B_{\alpha \gamma}\equiv\partial_\gamma  \beta_\alpha(u^*)$:
\begin{eqnarray}
\label{reu:H2}
k\,\partial_k\,{u}_\alpha(k)=\sum\limits_\gamma B_{\alpha \gamma}\,\left(u_\gamma(k)
-u_{\gamma}^*\right)\;.
\end{eqnarray}
The general solution to this equation reads
\begin{eqnarray}
\label{reu:H3}
u_\alpha(k)=u_{\alpha}^*+\sum\limits_I C_I\,V^I_\alpha\,
\left(\frac{k_0}{k}\right)^{\theta_I}
\end{eqnarray}
where the $V^I$'s are the right-eigenvectors of ${\bf B}$ with eigenvalues 
$-\theta_I$, i.e., $\sum_{\gamma} B_{\alpha \gamma}\,V^I_\gamma =-\theta_I\,V^I_\alpha$. Since ${\bf B}$ is not symmetric in general the $\theta_I$'s are not guaranteed to be real. We
assume that the eigenvectors form a complete system though. Furthermore, $k_0$ 
is a fixed reference scale, and the $C_I$'s are constants of integration. The quantities $\theta_I$ are referred to as \emph{critical exponents}
since when the renormalization group is applied to critical 
phenomena (second order phase transitions) the traditionally 
defined critical exponents are related to the $\theta_I$'s in a 
simple way \cite{reu:avactrev}.

If $u_\alpha(k)$ is to describe a trajectory in $\cS_{\rm UV}$,
 $u_\alpha(k)$ must 
approach $u_{\alpha}^*$ in the limit
$k\rightarrow\infty$ and therefore we must set $C_I=0$ for all $I$ with 
${\rm Re}\,\theta_I<0$. Hence the dimensionality $\Delta_{\rm UV}$ equals the 
number of ${\bf B}$-eigenvalues with a negative real part, i.e., the number of
$\theta_I$'s with ${\rm Re}\,\theta_I >0$. The corresponding eigenvectors 
span the tangent space to
$\cS_{\rm UV}$ at the NGFP.
If we {\it lower} the cutoff for a generic trajectory with all
$C_I$ nonzero, only $\Delta_{\rm UV}$ 
``relevant'' parameters
corresponding to the eigendirections tangent to $\cS_{\rm UV}$ grow 
(${\rm Re}\, \theta_I > 0$), while the remaining ``irrelevant'' couplings 
pertaining to the eigendirections normal to $\cS_{\rm UV}$ decrease 
(${\rm Re}\, \theta_I < 0$). Thus near the NGFP a generic trajectory 
is attracted towards $\cS_{\rm UV}$.

Coming back to the Asymptotic Safety construction, let us now 
use this fixed point in order to take the limit $k \ra \infty$. 
The trajectories which define an infinite cutoff limit are 
special in the sense that all irrelevant couplings are set to zero: $C_I = 0$ 
if ${\rm Re} \, \theta_I < 0$. These conditions place the trajectory 
exactly on $\cS_{\rm UV}$. There is a $\Delta_{\rm UV}$-parameter family 
of such trajectories, and the experiment must decide which one is 
realized in Nature. 
Therefore the predictive power of the theory increases with decreasing
 dimensionality of ${\cal S}_{\rm UV}$, i.e., number of UV attractive eigendirections of the
NGFP. If $\Delta_{\rm UV} < \infty$, the quantum field 
theory thus constructed is comparable to and as predictive as a perturbatively
renormalizable model with $\Delta_{\rm UV}$ ``renormalizable couplings''. 
Summarizing, we call a theory ``Asymptotically Safe'', if its UV behavior is controlled by a non-Gaussian fixed point with a finite number of relevant directions. 
The former condition ensures that the theory is safe from unphysical UV divergences while the latter requirement guarantees the predictivity of the construction. 
\index{asymptotically safe theory}

Up to this point our discussion did not involve any approximation. A method
which gives rise to non-perturbative approximate solutions is to truncate the
theory space $\{A[\,\cdot\,]\}$. The basic idea is to project the RG flow onto
a finite dimensional subspace of theory space. The subspace should be chosen
in such a way  that the projected flow encapsulates the essential physical 
features of the exact flow on the full space.

Concretely the projection onto a truncation\index{truncation} subspace is performed 
as follows. One makes an ansatz of the form 
$ 
\Gamma_k[\Phi, \bar{\Phi}] = \sum_{i=1}^N {\ub}_i(k) P_i[\Phi, \bar{\Phi}]\,,
$
where the $k$-independent functionals
$\{P_i[\, \cdot \,], i=1,\cdots,N \}$ form a `basis' on the subspace selected. 
For a scalar field $\phi$, say, examples include pure potential terms 
$\int d^dx \phi^m(x)$, 
$\int d^dx \phi^n(x) \ln \phi^2(x)$, $\cdots$, a standard kinetic 
term $\int \! d^dx (\dd \phi)^2$, higher order derivative terms 
$\int \! d^dx \, \phi \left({\dd^2} \right)^n \phi$, 
 $\cdots$, and non-local terms like 
$\int \!d^dx \, \phi \ln(-\dd^2) \phi$, $\cdots$.   
Even if $\Gamma_{\infty}$ is simple, a standard $\phi^4$ action,
say, the evolution from $k =\infty$ downwards will generate such     
terms.

The projected RG flow is described by a set of ordinary (if $N < \infty$) 
differential equations for the couplings $\ub_i(k)$. They arise as follows.
 Let us assume we 
expand the $\Phi$-dependence of $\frac{1}{2}{\rm Tr}[\cdots]$ 
(with the ansatz for $\Gamma_k[\Phi, \bar{\Phi}]$ inserted) in a basis
$\{P_{\alpha}[\, \cdot \,]\}$ of the {\it full} theory space which contains  
the $P_i$'s spanning the truncated space as a subset: 
\be 
\frac{1}{2} {\rm Tr}[\cdots] =  
\sum_{\alpha =1}^{\infty} \overline{\beta}_{\alpha}(\ub_1, \cdots, \ub_N;k) 
\, P_{\alpha}[\Phi, \bar{\Phi}]     
=      
\sum_{i =1}^N \overline{\beta}_i(\ub_1, \cdots, \ub_N;k) 
\, P_i[\Phi, \bar{\Phi}] + {\rm rest}\,. 
\label{reu:E15}
\ee
Here the ``rest'' contains all terms outside the truncated theory 
space; the approximation consists in neglecting precisely
those terms. Thus, equating (\ref{reu:E15}) to the LHS of the flow equation,      
$\dd_t \Gamma_k = \sum_{i=1}^N \dd_t \ub_i(k) P_i$, the linear independence 
of the $P_i$'s implies the coupled system of ordinary differential 
equations 
\be 
\dd_t \ub_i(k) = \overline{\beta}_i(\ub_1,\cdots , \ub_N;k)\,,
\quad i = 1, \cdots, N\,.
\label{reu:E16}
\ee
Solving (\ref{reu:E16}) one obtains an {\it approximation} to the 
exact RG trajectory\index{RG trajectory} projected onto the chosen subspace. Note that 
this approximate trajectory does, in general, not coincide with 
the projection of the exact trajectory, but if the subspace 
is well chosen, it will not be very different from it.

\section{The effective average action for gravity}
\label{reu:sect:3}
The effective average action\index{effective average action!for gravity|textbf} for gravity which has been introduced in
ref.\ \cite{reu:mr} is a concrete implementation of the general ideas outlined above.
The ultimate goal is to give meaning to an integral over `all' 
metrics $\gamma_{\mu\nu}$ of the form $\int \! \cD \gamma_{\mu\nu} \,
\exp\{ - S[\gamma_{\mu\nu}] + {\rm source \; terms}\}$ whose 
bare action $S[\gamma_{\mu\nu}]$\index{bare action} is invariant under general coordinate transformations.
The first step consists in splitting the quantum metric according to
\be\label{reu:backsplit}
\gamma_{\mu\nu} = \gb_{\mu\nu} + h_{\mu\nu}
\ee
where $\gb_{\mu\nu}$ is a fixed, but unspecified, background metric\index{background field}\index{background metric} and 
$ h_{\mu\nu}$ are the quantum fluctuations\index{quantum metric fluctuations} around this background which are not necessarily
small. This allows the formal construction of the gauge-fixed (Euclidean) gravitational path integral
\index{path integral, gravitational}
\be\label{reu:PI1}
\int \cD h \cD C^\mu \cD \bar{C}_\mu \exp\{ - S[\gb+h] - S^{\rm gf}[h; \gb] - S^{\rm ghost}[h, C, \bar{C}; \gb] - \Delta_k S[h, C, \bar{C}; \gb] \} \, .
\ee
Here $S[\gb+h]$ is a generic action, which depends on $\gamma_{\mu\nu}$ only, while the background gauge fixing $S^{\rm gf}[h; \gb]$\index{gauge fixing}\index{gauge fixing!gauge fixing action} and ghost contribution $S^{\rm ghost}[h, C, \bar{C}; \gb]$
\index{ghost action} contain $\gb_{\mu\nu}$ and  $h_{\mu\nu}$ in such a way that they do not combine into a full
$\gamma_{\mu\nu}$. We take $S^{\rm gf}[h; \gb]$ to be a gauge fixing ``of the background type'' \cite{reu:dewitt-books},
i.e. it is invariant under diffeomorphisms acting on both $h_{\mu\nu}$ and $\bar{g}_{\mu\nu}$. 

The key ingredient in the construction of the FRGE is the coarse graining term
$\Delta_k S[h,C,\bar{C}; \gb]$. It is quadratic in the fluctuation field,
$\int d^dx \sqrt{\gb} \, h_{\mu\nu} \cR^{\mu\nu\rho\sigma}_k(-\bar{D}^2) h_{\rho\sigma}\,$,
plus a  similar term  for the ghosts. 
The kernel $\cR^{\mu\nu\rho\sigma}_k(p^2)$ provides a $k$-dependent mass term which separates the fluctuations into high momentum modes $p^2 \gg k^2$ and low momentum modes $p^2 \ll k^2$ with respect to the scale set by the covariant Laplacian of the background metric. 
The profile  of  $\cR^{\mu\nu\rho\sigma}_k(p^2)$ ensures that the high momentum modes are integrated out unsuppressed while the contribution of the low momentum modes to the path integral is suppressed by the $k$-dependent  mass term. Varying $k$ then naturally realizes Wilson's idea of coarse graining by integrating out the quantum fluctuations shell by shell.

The $k$-derivative of eq.\ \eqref{reu:PI1} with $h_{\mu\nu}$ and the ghosts coupled to appropriate sources, provides the
starting point for the construction of the functional renormalization group equation\index{FRGE} for the effective average action $\Gamma_k$  \cite{reu:avact,reu:ym}.(See \cite{reu:avactrev}
for reviews.) For gravity this flow equation\index{flow equation} takes the form \cite{reu:mr}
\be
\p_t \Gamma_k[\bar{h}, \xi, \bar{\xi}; \gb] = \half {\rm STr} \left[ \left( \Gamma_k^{(2)} + \cR_k \right)^{-1} \, \p_t \cR_k  \right]\, .
\label{reu:FRGEgeneral}
\ee
Here $t = \log(k/k_0)$, {\rm STr} is a  functional supertrace which includes a minus sign for the ghosts $\xi\equiv\langle C\,\rangle, \bar{\xi}\equiv\langle \bar{C}\,\rangle$, $\cR_k$ is the matrix valued (in field space) IR cutoff
\index{cutoff (infrared)} introduced above, and $\Gamma_k^{(2)}$ is the second variation of $\Gamma_k$ with respect to the
{\it fluctuation fields}. Notably, $\Gamma_k[\bar{h}, \xi, \bar{\xi}; \gb]$ depends on {\it two} metrics, $\gb_{\mu\nu}$ and
\be
g_{\alpha\beta}  \equiv \langle \gamma_{\alpha\beta} \rangle = \gb_{\alpha\beta} + \bar{h}_{\alpha\beta}  \, , \qquad  \bar{h}_{\alpha\beta} \equiv \langle h_{\alpha\beta} \rangle \, .
\ee
In this sense, $\Gamma_k$ is of an intrinsically {\it bimetric} nature, and therefore we often write
$\Gamma_k[g, \gb, \xi, \bar{\xi}] \equiv \Gamma_k[\bar{h} = g - \gb, \xi, \bar{\xi}; \gb]$.
This functional is invariant under background gauge transformations\index{background gauge transformations} acting on all
four fields simultaneously.
It is a $k$-dependent generalization of the standard effective action $\Gamma\equiv\Gamma_0$\index{effective action} to which
it reduces in the limit $k\rightarrow 0$. It can also be shown that $\Gamma_k$ in the limit $k\rightarrow\infty$ is essentially
equivalent to the bare action $S$. (For further details about $\Gamma_k$ for gravity we refer to \cite{reu:mr}.)

\section{The Einstein-Hilbert truncation}
\label{reu:sect:4}
\index{truncation}\index{Einstein-Hilbert truncation|textbf}
Solving the FRGE (\ref{reu:FRGEgeneral}) is equivalent to (and as difficult as) calculating
the functional integral over $\gamma_\mn$. It is therefore important 
to devise efficient approximation methods. The truncation of theory 
space is the one which makes maximum use of the FRGE reformulation of the 
quantum field theory problem at hand. 

The first truncation for which the RG flow has been worked out \cite{reu:mr} is the ``Einstein-Hilbert
truncation'' which retains in $\Gamma_k$ only the terms
$\int\! d^dx \, \sqrt{g}$ and $\int\! d^dx \, \sqrt{g} R$, already present in the  
in the classical action, with $k$-dependent coupling constants,
as well as the classical gauge fixing and ghost terms:\index{Einstein-Hilbert action}
\be
\label{reu:G3}
\Gamma_k = \frac{1}{16\pi\,G_k}\int\! d^dx \,\sqrt{g} \left\{ 
-R + 2\bar\lambda_k \right\}
+ \;\text{class. gf- and gh-terms}\,.
\ee
In this case the truncation subspace is 2-dimensional. The ansatz 
(\ref{reu:G3}) contains two free functions of the scale, the running 
cosmological constant $\bar{\lb}_k$\index{cosmological constant|textbf}\index{cosmological constant!running} and the running
Newton constant $G_k$\index{Newton's constant|textbf}\index{Newton's constant!running}.

Upon inserting the ansatz (\ref{reu:G3}) into the flow 
equation (\ref{reu:FRGEgeneral}) it boils down to a system of two ordinary differential 
equations. We shall display them here in terms of the 
dimensionless running cosmological constant\index{cosmological constant!dimensionless} and Newton constant
\index{Newton's constant!dimensionless}, respectively:
\be \label{reu:G11}
\lb_k \equiv k^{-2} \bar{\lb}_k\,, \qquad
 g_k \equiv k^{d-2} G_k \,.
\ee  
Using $\lambda_k$ and $g_k$ the RG equations become autonomous
\be\label{reu:EHflow}
k \p_k g(k) = \beta_g( g(k), \lambda(k)) \, , \qquad k \p_k \lambda(k) = \beta_\lambda(g(k), \lambda(k)) \, ,
\ee
with
\be
\beta_g(g_k, \lb_k) = \big[d - 2 + \eta_N(g_k, \lambda_k) \big]\, g_k \, .
\ee
Here $\eta_N \equiv \partial_t \ln G_k$ is the anomalous dimension\index{anomalous dimension|textbf} of the
operator $\sqrt{g} R$. The explicit form of the beta functions $\beta_g$
and $\beta_\lambda$ for arbitrary cutoff $\cR_k$ and dimension can be found in ref.\ \cite{reu:mr}.
 Here we only display the result for $d=4$ and
a sharp cutoff\index{cutoff (infrared)}:
\begin{subeqnarray} 
\dd_t \lb_k \is -(2 - \eta_N) \lb_k - \frac{g_k}{\pi} 
\Big[ 5\ln(1 - 2 \lb_k) - 2 \zeta(3) + \frac{5}{2} \eta_N\Big]\,,
\\
\dd_t  g_k \is (2 + \eta_N) \,  g_k\,,
\\
\eta_N \is - \frac{2 \, g_k}{ 6\pi + 5 \, g_k} 
\Big[ \frac{18}{1 - 2 \lb_k} +  5\ln(1 - 2 \lb_k) - 
\zeta(2) + 6 \Big]\,.
\label{reu:G18}
\end{subeqnarray}

\begin{figure}[t]
\centering
\includegraphics[width=.95\columnwidth]{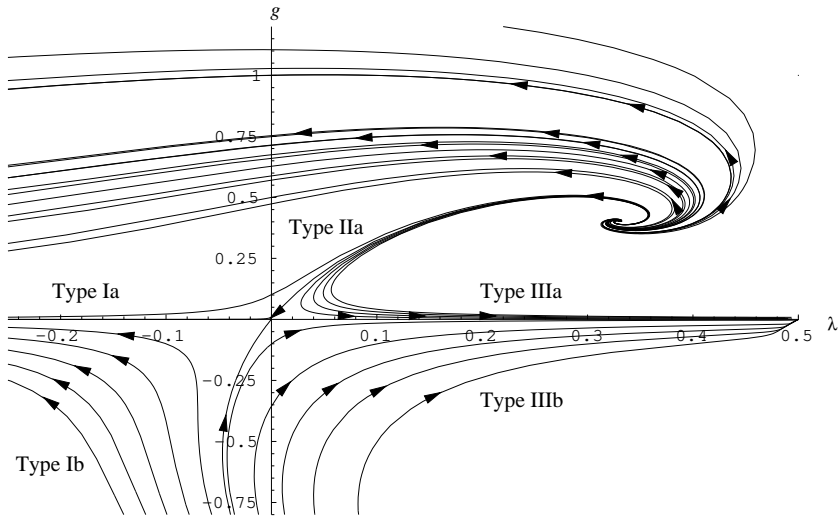}
\caption{RG flow in the $g$-$\lb-$plane. The arrows point in the direction 
of increasing coarse graining, i.e., of decreasing $k$. (From \cite{reu:frank1}.)}
\label{reu:fig0}
\end{figure}

In \cite{reu:frank1} this system has  
 been analyzed in detail, using both 
analytical and numerical methods. In particular all RG trajectories
 have been classified, and examples
have been computed numerically. The most important classes of 
trajectories in the phase portrait\index{phase portrait (flow diagram)!of the Einstein-Hilbert truncation} on the
$g$-$\lb-$plane are shown in Fig.~\ref{reu:fig0}.

The RG flow is found to be dominated by two fixed points $(g^*, \lb^*)$:
the GFP\index{Gaussian fixed point (GFP)} at $g^* = \lb^* =0$, and 
a NGFP with $g^* > 0$ and 
$\lb^* >0$. There are three classes of trajectories emanating from the NGFP:
trajectories of Type Ia and IIIa run towards negative and positive
cosmological constants, respectively, and the single trajectory of
Type IIa (``separatrix'') hits the GFP for $k\to 0$. The
high momentum properties of QEG are governed by the 
NGFP; for $k \to \infty$, in Fig.\ \ref{reu:fig0} 
all RG trajectories on the half--plane
$g>0$ run into this point. The two critical exponents are a complex conjugate pair
$\theta_{1,2}=\theta'\pm\I\theta''$ with $\theta'>0$.
The fact that at the NGFP the dimensionless coupling constants $g_k, \lambda_k$ approach constant, non-zero values then implies that the dimensionful quantities
run according to
\be
\label{reu:asymrun}
G_k = g^* k^{2-d} \;, \qquad \bar{\lambda}_k = \lambda^*\,k^2 \, .
\ee 
Hence for $k \ra \infty$ and $d > 2$ the dimensionful Newton constant\index{Newton's constant} vanishes
while the cosmological constant\index{cosmological constant} diverges.

So, the Einstein-Hilbert truncation\index{Einstein-Hilbert truncation} does indeed predict the existence of a NGFP with exactly
the properties needed for the Asymptotic Safety\index{Asymptotic Safety} construction. Clearly the crucial question is whether
the NGFP found is the projection of an exact fixed point in the full theory or 
merely the artifact of an insufficient approximation. This question has been analyzed during the past decade within truncations
of ever increasing complexity. All investigations performed to date support the existence of a NGFP\index{NGFP} in the exact
theory, and without exception they predict a projected RG flow on the $g$-$\lambda$--plane which is qualitatively similar to
that of the Einstein-Hilbert truncation. In fact, the phase portrait in Fig.\ \ref{reu:fig0} has survived substantial
generalizations of the truncation ansatz for the average action. Furthermore, clear evidence for a small, finite
dimensionality of $\cS_{\rm UV}$ was found, first in $2+\epsilon$ dimensions \cite{reu:oliver2} and then by an
impressively complex calculation in $d=4$ also \cite{reu:Codello:2008vh,reu:MS1}.

Besides its successes in describing gravity at high energies, QEG also recovers classical general relativity at low energies.
Concretely, it was shown in \cite{reu:bh} that Fig.\ \ref{reu:fig0} contains Type IIIa trajectories
which are in agreement with observational data. This analysis is fairly robust 
and clear-cut; it does not involve the NGFP. All that is needed is the RG flow linearized 
about the GFP. In its vicinity one
has \cite{reu:mr}
\be\label{reu:linflow}
\lbar(k) = \lbar_0 + \nu \, \barg \, k^d + \cdots  \, , \qquad  G(k) =\barg+\cdots \, ,
\ee
i.e., $\lbar$ displays a running $\propto k^d$
and $G$ is approximately constant. Here $\nu$ is a positive 
constant of order unity \cite{reu:mr,reu:frank1}.
These equations are valid if $\lambda(k) \ll 1$ and $g(k)\ll 1$. They
describe a 2-parameter family of RG trajectories labeled by the pair $(\lbar_0, \barg)$.
It will prove convenient to use an alternative labeling $(\lat, \kat)$ with
$\lat \equiv (4  \nu  \lbar_0 \barg)^{1/2}$ and 
$\kat \equiv   ( {\lbar_0}/{\nu  \barg}  )^{1/4} $.
The old labels are expressed in terms of the new ones as
$\lbar_0 = \frac{1}{2} \lat \; \kat^2$ and $\barg = {\lat}/(2\nu\,\kat^2)$.
It is furthermore convenient to introduce the abbreviation
$\gat\equiv {\lat}/(2\nu)$.
\begin{figure}[t]
\begin{center}
\includegraphics[width=.95\columnwidth]{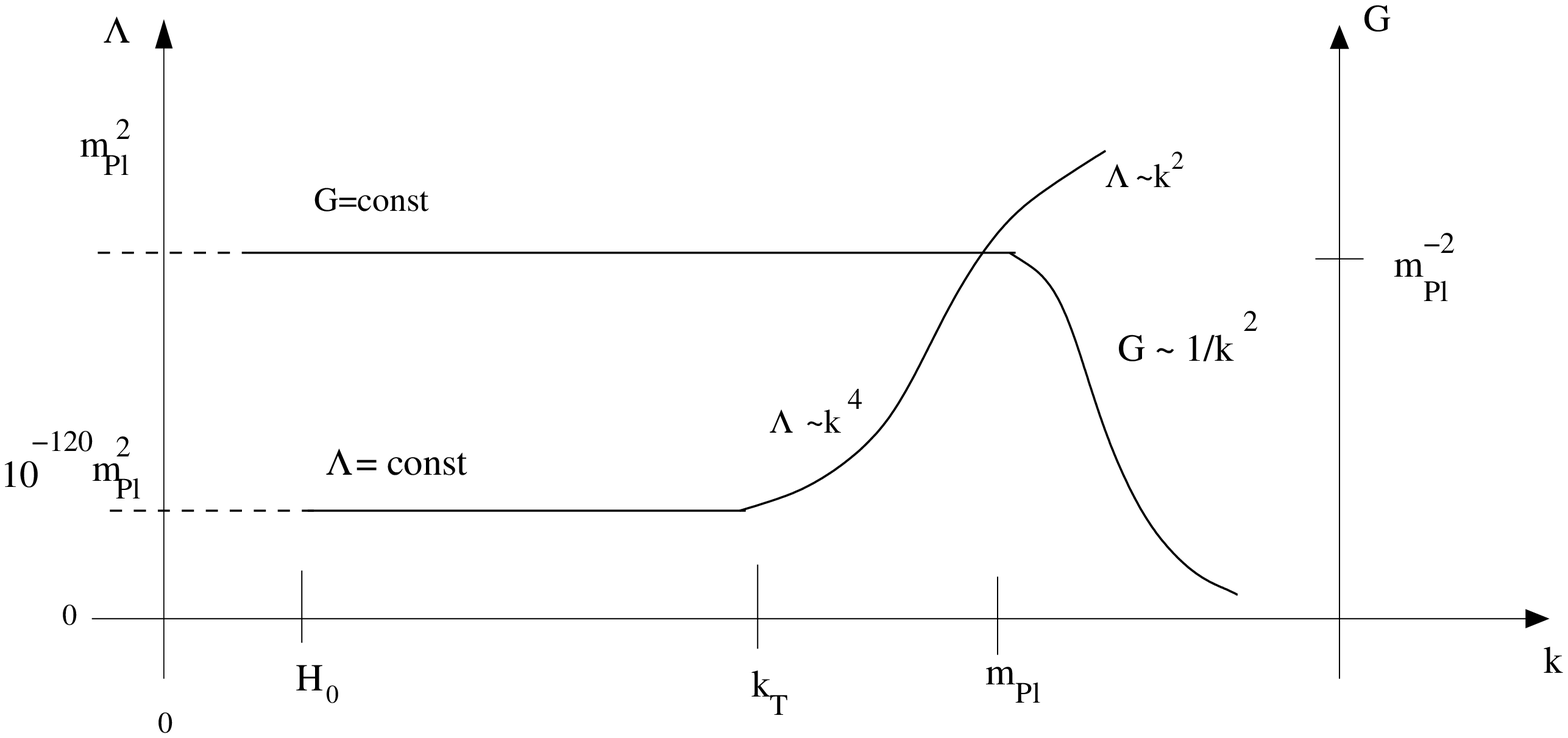}
\end{center}
\caption{The dimensionful $\Lambda(k)\equiv\lbar(k)$ and $G(k)$ for a Type IIIa trajectory with
realistic parameters.}
\label{reu:figGL}
\end{figure} 
When parameterized by the pair $(\lat,\kat)$ the trajectories assume the form
\ba\label{reu:1.16}
&&\lbar(k) = \frac{1}{2} \; \lat\; \kat^2 \; \Big [ 1+(k/\kat)^4\Big ]
\equiv \lbar_0 \Big [ 1+(k/\kat)^4 \Big ]\,,\\[2mm]
&&G(k) = \frac{\lat}{2\,\nu\,\kat^2}\equiv \frac{\gat}{\kat^2}\;,\nonumber
\ea
or, in dimensionless form, 
\be\label{reu:1.17}
\lambda(k) = \frac{1}{2} \; \lat \Big  [  \Big (\frac{\kat}{k} \Big)^2+ 
 \Big ( \frac{k}{\kat}  \Big)^2 \Big ], \;\;\;\;\;\;\;\;\;\;  g(k) = \gat \,  \Big( \frac{k}{\kat}\Big  )^2\,.
\ee
As for the interpretation of the new variables, 
it is clear that $\lat \equiv \lambda(k\equiv \kat)$ and $\gat\equiv g(k=\kat)$,
while $\kat$ is the scale at which $\beta_\lambda$ (but not $\beta_g$) vanishes according
to the linearized running:
$\beta_\lambda(\kat)\equiv k{d \lambda(k)}/{dk}  |_{k=\kat} =0$.
Thus we see that $(\gat,\lat)$ are the coordinates of the turning point T
of the Type IIIa trajectory considered, and $\kat$ is the scale at which it is
passed. The regimes $k>\kat$ ($k<\kat)$ are conveniently referred to 
the ``UV regime'' (``IR regime''). 

Let us now hypothesize that, within a certain range of $k$-values, the RG trajectory
realized in Nature can be approximated by (\ref{reu:1.17}). In order to determine
its parameters $(\lbar_0, \barg)$ or $(\lat, \kat)$ we must perform a measurement
of $G$ and $\lbar$. If we interpret the observed values
$G_{\rm observed} = \mp^{-2}$, $\mp\approx 1.2\times 10^{19} \, {\rm GeV}$, and 
$\lbar_{\rm observed} \approx 10^{-120}\, \mp^2\nonumber$
as the running $G(k)$ and $\lbar(k)$ evaluated at a scale $k\ll \kat$, then we get from
(\ref{reu:1.16}) that $\lbar_0 =\lbar_{\rm observed} $ and 
$\barg = G_{\rm observed}$. Using the definitions of  $\lat$ and $\kat$ along with $\nu = O(1)$ this leads to the 
order-of-magnitude estimates
$\gat\approx \lat \approx 10^{-60}$ and $\kat\approx 10^{-30}\;\mp\approx (10^{-3} {\rm cm})^{-1}$.
Because of the tiny values of $\gat$ and $\lat$ the turning point lies in the linear regime of the GFP. 
Going beyond the linear regime, the $k$-dependence of $G$ and $\lbar$ is plotted schematically
in Fig.~\ref{reu:figGL}.

\section{The multifractal properties of QEG space-times}
\label{reu:sect:7}
We now proceed by discussing an intriguing consequence arising from the scale-dependence of the
gravitational effective action, namely that the QEG\index{QEG} space-time at short distances develops
fractal\index{fractals} properties\index{space-time!multifractal properties}\index{multifractal space-time|see{space-time}} \cite{reu:oliver1,reu:oliver2,reu:oliverfrac}. As we have seen, the effective average action $\Gamma_k[g_{\mu\nu}]$ defines
an infinite set of effective field theories, valid near a variable mass scale $k$. Intuitively speaking,
the solution $\big<g_{\mu\nu}\big>_k$ of the scale dependent field equation\index{effective field equations}
\begin{eqnarray}
\label{reu:fe}
\frac{\delta\Gamma_k}{\delta g_{\mu\nu}(x)}[\big<g\big>_k]=0
\end{eqnarray}
can be interpreted as the metric averaged over (Euclidean) space-time volumes
of a linear extension $\ell$ which typically is of the order of $1/k$. Knowing
the scale dependence of $\Gamma_k$, i.e., the renormalization group trajectory
$k\mapsto\Gamma_k$, we can in principle follow the solution 
$\big<g_{\mu\nu}\big>_k$ from the ultraviolet $(k\rightarrow\infty)$ to the
infrared $(k\rightarrow 0)$.
\smallskip

\noindent
\textbf{(1) ``Quantum space-times''.}
It is an important feature of this approach that the infinitely many equations
of (\ref{reu:fe}), one for each scale $k$, are valid {\it simultaneously}. They
all refer {\it to the same} physical system, the ``quantum space-time'', but
describe its effective metric structure on different scales. An observer using
a ``microscope'' with a resolving power $\ell \approx k^{-1}$ will perceive the Universe
to be a Riemannian manifold with metric $\big<g_{\mu\nu}\big>_k$.\footnote{The ``resolving power'' $\ell$ of the microscope\index{resolving power of a microscope} is in
general a complicated function of $k$. It can be found by an algorithm outlined
in \cite{reu:oliverfrac}. For the purposes of the present discussion it is
sufficient to think of this relationship as $\ell\approx 1/k$, like on flat space.} At every 
fixed $k$, $\big<g_{\mu\nu}\big>_k$ is a smooth classical metric. But since
the quantum space-time is characterized by the infinity of equations (\ref{reu:fe})
with $k=0,\cdots,\infty$ it can acquire very non-classical and in particular
fractal\index{space-time!multifractal properties} features.
In particular, it was concluded in 
\cite{reu:oliver1,reu:oliver2} that the effective dimensionality of space-time is
scale dependent. It equals 4 at macroscopic distances ($\ell\gg\ell_{\rm Pl}$)
but, near $\ell\approx\ell_{\rm Pl}$, it gets dynamically reduced to the value
2. For $\ell\ll\ell_{\rm Pl}$ space-time resembles a 2-dimensional fractal. In the
following we review the arguments that led to this conclusion.
\smallskip

\noindent
\textbf{(2) Self-similarity in the fixed point regime.}\index{self-similarity of space-time}\index{space-time!self-similarity}
For simplicity we use the Einstein-Hilbert truncation to start with, and we
consider space-times with classical dimensionality $d=4$. The corresponding RG
trajectories are shown in Fig.\ \ref{reu:fig0}. The physically relevant ones, for $k\rightarrow\infty$, all 
approach the NGFP at $(g_*,\lambda_*)$ so that the dimensionful quantities
run according to \eqref{reu:asymrun}. 
This scaling behavior is realized in the asymptotic scaling regime
$k\gg m_{\rm Pl}$. Near $k= m_{\rm Pl}$ the trajectories cross over towards
the GFP at $g=\lambda=0$, and then run towards negative,
vanishing, and positive values of $\lambda$, respectively. 
For our present purpose, it suffices to consider the limiting cases of very small
and very large distances of a RG trajectory. We assume that $G_k$ and $\bar{\lambda}_k$ behave as in 
(\ref{reu:asymrun}) for $k\gg m_{\rm Pl}$, and that they are constant for
$k\ll m_{\rm Pl}$. The precise interpolation between the two regimes will not be needed here.

The argument of ref. \cite{reu:oliver2} concerning the fractal\index{space-time!multifractal properties} nature of the QEG
space-times was as follows. Within the Einstein-Hilbert truncation of theory
space, the effective field equations (\ref{reu:fe}) happen to coincide with the 
ordinary Einstein equation, but with $G_k$ and $\bar{\lambda}_k$ replacing the
classical constants. Without matter,
\be\label{reu:effeom}
R_{\mu\nu}(\langle g \rangle_k) = \frac{2}{2-d} \, \bar{\lambda}_k \,  \langle g_{\mu\nu} \rangle_k \, .
\ee
Since in absence of dimensionful constants of integration
$\bar{\lambda}_k$ is the only quantity in this equation which sets a 
scale, every solution to (\ref{reu:effeom}) has a typical radius of curvature 
$r_c(k)\propto 1/\sqrt{\bar{\lambda}_k}$. (For instance, the maximally 
symmetric $S^4$-solution has the radius $r_c=r=\sqrt{3/\bar{\lambda}_k}$.) 
If we want to explore the space-time structure at a fixed length scale $\ell$ 
we should use the action $\Gamma_k[g_{\mu\nu}]$ at $k=1/\ell$ because with 
this functional a tree level analysis is sufficient to describe the essential 
physics at this scale, including the relevant quantum effects. Hence, when we 
observe the space-time with a microscope of resolution $\ell$, we will see an 
average radius of curvature given by 
$r_c(\ell)\equiv r_c(k=1/\ell)$. Once $\ell$ is
smaller than the Planck length $\ell_{\rm Pl}\equiv m_{\rm Pl}^{-1}$\index{Planck scale}
we are in the fixed point regime where $\bar{\lambda}_k\propto k^2$ so that 
$r_c(k)\propto 1/k$, or
\be
\label{reu:radius}
\framebox{\; $r_c(\ell)\propto\ell$ \big. \;} \; .
\ee
Thus, when we look at the structure of space-time with a microscope of 
resolution $\ell\ll\ell_{\rm Pl}$, the average radius 
of curvature which we measure is proportional to the resolution 
itself. If we want to probe finer details and decrease $\ell$ we automatically
decrease $r_c$ and hence {\it in}crease the average curvature. Space-time seems
to be more strongly curved at small distances than at larger ones. The 
scale-free relation (\ref{reu:radius}) suggests that at distances below the Planck
length the QEG\index{QEG} space-time is a special kind of fractal with a self-similar 
structure. It has no intrinsic scale because in the fractal regime, i.e., when 
the RG trajectory is still close to the NGFP\index{NGFP}, the parameters which usually
set the scales of the gravitational interaction, $G$ and $\bar{\lambda}$, are 
not yet ``frozen out''. This happens only later on, somewhere half way between
the non-Gaussian and the Gaussian fixed point, at a scale of the order of 
$m_{\rm Pl}$.

Below this scale, $G_k $ and $\bar{\lambda}_k$ stop running and, as a result,
$r_c(k)$ becomes independent of $k$ so that $r_c(\ell)={\rm const}$ for 
$\ell\gg\ell_{\rm Pl}$. In this regime $\big<g_{\mu\nu}\big>_k$ is 
$k$-independent, indicating that the macroscopic space-time is describable by a
single smooth, classical Riemannian manifold.
\smallskip

\noindent
\textbf{(3) Anomalous dimension and graviton propagator.}\index{graviton propagator}\index{anomalous dimension}
An independent argument supporting the assertion that the QEG space-time has an effective
dimensionality which is $k$-dependent and noninteger in general based upon the .
\emph{anomalous dimension} $\eta_N\equiv \partial_t \ln G_k$ has been put forward in 
ref. \cite{reu:oliver1}.
In a sense which we shall make more precise in a
moment, the effective dimensionality of space-time equals $4+\eta_N$. The RG
trajectories of the Einstein-Hilbert truncation (within its domain of validity)
have $\eta_N\approx 0$ for $k\rightarrow 0$ and $\eta_N\approx -2$ for 
$k\rightarrow\infty$, the smooth change by two units occuring near 
$k\approx m_{\rm Pl}$. As a consequence, the effective dimensionality is 4 for
$\ell\gg\ell_{\rm Pl}$ and 2 for $\ell\ll\ell_{\rm Pl}$.

In fact, the UV fixed point has an anomalous dimension $\eta\equiv\eta_N(g_*,
\lambda_*)=-2$. We can use this information in order to determine the 
momentum dependence of the dressed graviton propagator for momenta $p^2\gg
m_{\rm Pl}^2$. Expanding \eqref{reu:G3}
about flat space
and omitting the standard tensor structures we find the inverse propagator
$\widetilde{\cal G}_k(p)^{-1}\propto G_k^{-1}\,p^2$. The conventional dressed 
propagator $\widetilde{\cal G}(p)$ contained in $\Gamma\equiv\Gamma_{k=0}$ 
is obtained from the {\it exact} $\widetilde{\cal G}_k$ in the limit 
$k\rightarrow 0$.
For $p^2>k^2\gg m_{\rm Pl}^2$ the actual cutoff scale is the physical momentum
$p^2$ itself so that the $k$-evolution of $\widetilde{\cal G}_k(p)$ stops at the
threshold $k=\sqrt{p^2}$. Therefore
\begin{eqnarray}
\label{reu:gp1}
\widetilde{\cal G}(p)^{-1}\propto\;p^2\,G_k^{-1}\Big|_{k=\sqrt{p^2}}\;\propto\;
(p^2)^{1-\frac{\eta}{2}}
\end{eqnarray}
because $G_k^{-1}\propto k^{-\eta}$ when $\eta$ is 
(approximately) constant. In $d$ flat dimensions, and for $\eta\neq 2-d$, the
Fourier transform of $\widetilde{\cal G}(p)\propto 1/(p^2)^{1-\eta/2}$ yields 
the following propagator in position space:
\begin{eqnarray}
\label{reu:gp2}
{\cal G}(x;y)\propto\;\frac{1}{\left|x-y\right|^{d-2+\eta}}\;.
\end{eqnarray}
This form of the propagator is well known from the theory of critical 
phenomena, for instance. (In the latter case it applies to large distances.)
Eq. (\ref{reu:gp2}) is not valid directly at the NGFP. For $d=4$ and $\eta=-2$
the dressed propagator\index{graviton propagator} is $\widetilde{\cal G}(p)=1/p^4$ which has the following
representation in position space:
\begin{eqnarray}
\label{reu:gp3}
{\cal G}(x;y)=-\frac{1}{8\pi^2}\,\ln\left(\mu\left|x-y\right|\right)\;.
\end{eqnarray}
Here $\mu$ is an arbitrary constant with the dimension of a mass. Obviously
(\ref{reu:gp3}) has the same form as a $1/p^2$-propagator in 2 dimensions.

Slightly away from the NGFP, before other physical scales intervene, the 
propagator is of the familiar type (\ref{reu:gp2}) which shows that the quantity 
$\eta_N$
has the standard interpretation of an anomalous dimension in the sense that
fluctuation effects modify the decay properties of ${\cal G}$ so as to 
correspond to a space-time of effective dimensionality $4+\eta_N$.

Thus the properties of the RG trajectories imply a remarkable dimensional
reduction: \emph{Space-time, probed by a ``graviton'' with $p^2\ll m_{\rm Pl}^2$ is
4-dimensional, but it appears to be 2-dimensional for a graviton with 
$p^2\gg m_{\rm Pl}^2$} \cite{reu:oliver1}. More generally, 
 in $d$ classical dimensions, where the 
macroscopic space-time is $d$-dimensional, the anomalous dimension at the
fixed point is $\eta_N=2-d$. Therefore, for any $d$, the dimensionality of the
fractal as implied by $\eta_N$ is $d+\eta_N=2$ \cite{reu:oliver1,reu:oliver2}.

\section{Spectral, walk, and Hausdorff dimension}
\label{reu:sect:7c}
The fractal properties of the QEG space-time can be further quantified by 
investigating random walks and diffusion processes on fractals\index{space-time!multifractal properties}. In this course 
one is led to introduce various notions of fractal dimensions, such as the spectral or walk dimension \cite{reu:avra}.
A priori they have no reason to equal the effective dimension $d_{\rm eff}=d+\eta$ implied
by the running Newton constant and the graviton propagator.
\medskip

\noindent
{\bf (1) The spectral dimension.}\index{spectral dimension|textbf}
Consider the diffusion process where a spin-less test particle performs a Brownian random walk\index{random walk} on
an ordinary Riemannian manifold with a fixed classical metric $g_{\mu\nu}(x)$. It is described by the heat-kernel
\index{heat-kernel} $K_g(x, x^\prime; T)$ which gives the probability density for a transition of the particle from $x$ to
$x^\prime$ during the fictitious time $T$. It satisfies the heat equation
\be 
\p_T K_g(x, x^\prime; T) = - \Delta_g K_g(x, x^\prime; T) \, ,
\ee
where $\Delta_g = - D^2$ denotes the Laplace operator. In flat space, this equation is easily solved by
\be \label{reu:heat1}
K_g(x, x^\prime; T) = \int \frac{d^dp}{(2\pi)^d} \, \e^{i p \cdot (x-x^\prime)} \, \e^{-p^2 T} 
\ee
In general, the heat-kernel is a matrix element of the operator $\exp(- T \Delta_g)$. In the random walk picture
its trace per unit volume,
\be
P_g(T) = V^{-1} \int d^dx \sqrt{g(x)} \, K_g(x, x; T) \equiv V^{-1} \, \Tr \, \exp(- T \Delta_g) \, , 
\ee
has the interpretation of an average return probability. Here $V \equiv \int d^dx \sqrt{g(x)}$ denotes the total volume. It is well known that $P_g$ possesses an asymptotic early time expansion (for $T \rightarrow 0$) of the form $P_g(T) = (4 \pi T)^{-d/2} \sum_{n=0}^\infty A_n T^n$, with $A_n$ denoting the Seeley-DeWitt coefficients. From this expansion one can motivate the definition of the spectral dimension $d_s$ as the $T$-independent logarithmic derivative
\be \label{reu:DsTlim} 
d_s \equiv \left. - 2 \frac{d \ln P_g(T)}{d \ln T} \right|_{T = 0} \, . 
\ee
On smooth manifolds, where the early time expansion of $P_g(T)$ is valid, the spectral dimension agrees with the topological dimension $d$ of the manifold.

Given $P_g(T)$, it is natural to define an, in general $T$-dependent, generalization of the spectral dimension by
\be\label{reu:DsT}
\cD_s(T) \equiv - 2 \frac{d \ln P_g(T)}{d \ln T}\, .
\ee
 According to \eqref{reu:DsTlim}, we recover the true spectral dimension of the space-time by 
considering the shortest possible random walks, i.e., by taking the limit $d_s =  \lim_{T \rightarrow 0} \cD_s(T)$.
Note that in view of a possible comparison with other (discrete) approaches to quantum gravity
 the generalized, scale-dependent version \eqref{reu:DsT} will play a central role later on.
\medskip

\noindent
{\bf (2) The walk dimension.}\index{walk dimension|textbf}
Regular Brownian motion in flat space has the celebrated property
that the random walker's\index{random walk} average square displacement increases linearly
with time: $\langle r^2 \rangle \propto T$. Indeed, performing the integral
\eqref{reu:heat1} we obtain the familiar probability density
 \be \label{reu:FSHeat}
 K(x, x^\prime; T) = (4 \pi T)^{-d/2} \exp\left(- \frac{|x - x^\prime |^2}{4 T} \right)
 \ee
Using \eqref{reu:FSHeat} yields the expectation value $ \langle r^2 \rangle \equiv \langle x^2 \rangle = \int d^dx \, x^2 \, K(x, 0; T) \propto T$.

Many diffusion processes of physical interest (such as diffusion on fractals) are anomalous in the sense that this linear relationship is generalized to a power law
$\langle r^2 \rangle \propto T^{2/d_w}$ with $d_w \not = 2$. The interpretation of the so-called walk dimension $d_w$ is as follows. The trail left by the random walker is a random object, which is interesting in its own right. It has the properties of a fractal, even in the ``classical'' case when the walk takes place on a regular manifold. The quantity $d_w$ is precisely the fractal dimension of this trail. Diffusion processes are called regular if $d_w = 2$, and anomalous when $d_w \not = 2$.  
\medskip

\noindent
{\bf (3) The Hausdorff dimension.}\index{Hausdorff dimension|textbf}
Finally, we introduce the Hausdorff dimension $d_H$. Instead of working with its mathematically rigorous definition in terms of the Hausdorff measure and all possible covers of the metric space under consideration, the present, simplified definition may suffice for our present purposes. On a smooth set, the scaling law for the volume
$V(r)$ of a $d$-dimensional ball of radius $r$ takes the form 
\be\label{reu:Hdd}
V(r) \propto r^{d_H} \, .
\ee
The Hausdorff dimension is then obtained in the limit of infinitely small radius,
\be
d_H \equiv \lim_{r \rightarrow 0} \frac{\ln V(r)}{\ln r} \, .
\ee
 Contrary to the spectral or walk dimension whose definitions are linked to dynamical diffusion 
 processes on space-time, no such dynamics is associated with $d_H$.

\section{Fractal dimensions within QEG}
\label{reu:sect:7d}
Upon introducing various concepts for fractal dimensions in the last section, we now proceed with
their evaluation for the QEG effective space-times\index{space-time!multifractal properties}, following
refs.\ \cite{reu:oliverfrac} and \cite{reu:frankfrac}.
Our discussion will mostly be based on the Einstein-Hilbert truncation. As we shall see this restriction is actually
unnecessary in the asymptotic scaling regime, i.e., when the RG trajectory is close to the NGFP. In this case we can derive
{\it exact} results for the spectral and walk dimension by exploiting the scale invariance of the theory at the fixed point.
\medskip

\noindent
\textbf{(1) Diffusion processes on QEG space-times.}\index{diffusion processes on QEG space-times}
Since in QEG one integrates over all metrics, the central idea is to replace $P_g(T)$ by its expectation value
\be\label{reu:QPgT}
P(T) \equiv \langle P_\gamma(T) \rangle \equiv \int \cD \gamma \cD C \cD \bar{C} \;\; P_\gamma(T) \, \exp \left
(-S_{\rm bare}[\gamma, C, \bar{C}] \right ) \, .
\ee
Here $\gamma_{\m\nu}$ denotes the microscopic metric and $S_{\rm bare}$ is the bare action related to the UV fixed 
point, with the gauge-fixing and the pieces containing the ghosts $C$ and $\bar{C}$ included. For the untraced heat-kernel\index{heat-kernel}, we define likewise $K(x, x^\prime; T) \equiv \langle K_\gamma(x, x^\prime; T) \rangle \,$.
These expectation values are most conveniently calculated from the effective average action $\Gamma_k$, which equips 
the $d$-dimensional smooth manifolds underlying the QEG effective space-times with a family of metric structures $\left\{ \langle g_{\m\nu} \rangle_k, 0 \le k < \infty \right\}$, one for each coarse-graining scale $k$ \cite{reu:oliverfrac,reu:jan1}. These metrics are solutions to the effective field equations implied by $\Gamma_k$. 

We shall again approximate the latter by the Einstein-Hilbert truncation (\ref{reu:G3}).
The corresponding effective field equation\index{effective field equations} is given by \eqref{reu:effeom}. 
Based on this equation, we can easily
find the $k$-dependence of the corresponding solution $\langle g_{\m\nu} \rangle_k$ by rewriting \eqref{reu:effeom} as 
$[\bar{\lambda}_{k_0} / \bar{\lambda}_k] R^\m{}_\nu(\langle g \rangle_k) = \frac{2}{2-d} \bar{\lambda}_{k_0} \delta^\m{}_\nu$
for some fixed reference scale $k_0$, and exploiting that $R^\m{}_\nu(cg) = c^{-1} R^\m{}_\nu(g)$
for any constant $c > 0$. This shows that the metric and its inverse scale according to, for any $d$,
\be\label{reu:metricscaling}
\langle g_{\m\nu}(x) \rangle_k = [\bar{\lambda}_{k_0} / \bar{\lambda}_k] \langle g_{\m\nu}(x) \rangle_{k_0} \, , \qquad 
\langle g^{\m\nu}(x) \rangle_k = [\bar{\lambda}_k/ \bar{\lambda}_{k_0}] \langle g^{\m\nu}(x) \rangle_{k_0} \, .
\ee
Denoting the Laplace operators corresponding to the metrics $\langle g_{\m\nu} \rangle_k$ and $\langle g_{\m\nu} \rangle_{k_0}$ by $\Delta(k)$ and $\Delta(k_0)$, respectively, these relations imply
\be\label{reu:Laplacescaling}
\Delta(k) = \left[\bar{\lambda}_k / \bar{\lambda}_{k_0} \right] \Delta(k_0) \, .
\ee

At this stage, the following remark is in order. In the asymptotic scaling regime associated with the NGFP\index{NGFP} the scale-dependence of the couplings is fixed by the fixed point condition
\eqref{reu:asymrun}. This implies in particular
\be\label{reu:UVscaling}
\langle g_{\m\nu}(x) \rangle_k \propto k^{-2} \qquad (k \rightarrow \infty) \, .
\ee
This asymptotic relation is actually an {\it exact} consequence of Asymptotic Safety, which solely relies on the scale-independence of the theory at the fixed point.

We can evaluate the expectation value \eqref{reu:QPgT} by exploiting the effective field theory properties of the effective average action. Since $\Gamma_k$ defines an effective field theory at the scale $k$ we know that $\langle \cO(\gamma_{\m\nu}) \rangle \approx \cO(\langle g_{\m\nu} \rangle_k)$ provided the observable $\cO$ involves only momentum scales of the order of $k$. We apply this rule to the RHS of the diffusion equation, $\cO = - \Delta_\gamma K_\gamma(x, x^\prime; T)$. The subtle issue here is the correct identification of $k$. If the diffusion process involves (approximately) only a small interval of scales near $k$ over which $\bar{\lambda}_k$ does not change much, the corresponding heat equation contains the operator $\Delta(k)$ for this specific, fixed value of $k$: $\p_T K(x, x^\prime; T) = - \Delta(k) K(x, x^\prime; T)$. Denoting the eigenvalues of $\Delta(k_0)$ by $\cE_n$ and the corresponding eigenfunctions by $\phi_n$, this equation is solved by
\be\label{reu:heatk}
K(x, x^\prime; T) = \sum_n \phi_n(x) \phi_n(x^\prime) \exp\Big( - F(k^2) \cE_n T \Big) \, . 
\ee 
Here we introduced the convenient notation $F(k^2) \equiv \bar{\lambda}_k/\bar{\lambda}_{k_0}$. Knowing the propagation kernel, we can time-evolve any initial probability distribution $p(x; 0)$ according to 
\be
p(x; T) = \int d^dx^\prime \sqrt{g_0(x^\prime)} \, K(x, x^\prime; T) \, p(x^\prime; 0)
\ee
 with $g_0$ the determinant of $\langle g_{\m\nu} \rangle_{k_0}$. If the initial distribution has an eigenfunction expansion of the form $p(x; 0) = \sum_n C_n \phi_n(x)$ we obtain
\be\label{reu:prob2}
p(x; T) = \sum_n C_n \phi_n(x) \exp\Big( - F(k^2) \cE_n T \Big) \, .
\ee

If the $C_n$'s are significantly different from zero only for a single eigenvalue $\cE_N$, we are dealing with a single-scale problem and would identify $k^2 = \cE_N$ as the relevant scale at which the running couplings are to be evaluated. In general the $C_n$'s are different from zero over a wide range of eigenvalues. In this case we face a multiscale problem where different modes $\phi_n$ probe the space-time on different length scales. If $\Delta(k_0)$ corresponds to flat space, say, the eigenfunctions $\phi_n = \phi_p$ are plane waves with momentum $p^\m$, and they resolve structures on a length scale $\ell$ of order $1/|p|$. Hence, in terms of the eigenvalue $\cE_n \equiv \cE_p = p^2$ the resolution is $\ell \approx 1/\sqrt{\cE_n}$. This suggests that when the manifold is probed by a mode with eigenvalue $\cE_n$ it ``sees'' the metric $\langle g_{\m\nu} \rangle_k$ for the scale $k = \sqrt{\cE_n}$. Actually, the identification $k = \sqrt{\cE_n}$ is correct also for curved space since, in the construction of $\Gamma_k$, the parameter $k$ is introduced precisely as a cutoff in the spectrum of the covariant Laplacian.

As a consequence, under the spectral sum of \eqref{reu:prob2}, we must use the scale $k^2 = \cE_n$ which depends explicitly on the resolving power of the corresponding mode. Likewise, in eq.\ \eqref{reu:heatk}, $F(k^2)$ is to be interpreted as $F(\cE_n)$:
\be\label{reu:2.19}
\begin{split}
K(x, x^\prime; T) = & \, \sum_n \phi_n(x) \phi_n(x^\prime) \exp\Big(-F(\cE_n) \cE_n T \Big) \\
= & \, \sum_n \phi_n(x)  \exp\Big(-F\big(\Delta(k_0)\big) \Delta(k_0) T \Big) \phi_n(x^\prime) \, .
\end{split} 
\ee
As in \cite{reu:oliverfrac}, we choose $k_0$ as a macroscopic scale in the classical regime\index{space-time!classical regime}, and we assume that at $k_0$ the cosmological constant is small, so that
$\langle g_{\m\nu} \rangle_{k_0}$ can be approximated by the flat metric on $\mathbb{R}^d$. The eigenfunctions of
$\Delta(k_0)$ are plane waves then and eq.\ \eqref{reu:2.19} becomes
\be\label{reu:Heat:FlatSpace}
K(x, x^\prime; T) = \int \frac{d^dp}{(2 \pi)^d} \, e^{i p \cdot (x-x^\prime)} \, e^{-p^2 F(p^2) T}
\ee
where the scalar products are performed with respect to the flat metric, $\langle g_{\m\nu} \rangle_{k_0} = \delta_{\m\nu}$. The kernel \eqref{reu:Heat:FlatSpace} satisfies the relation $K(x, x^\prime; 0) = \delta^d(x-x^\prime)$ and, provided that
$\lim_{p \rightarrow 0} p^2 F(p^2) = 0$, also $\int d^dx K(x, x^\prime; T) = 1$. 

Taking the trace of \eqref{reu:Heat:FlatSpace} within this ``flat space-approximation'' yields \cite{reu:oliverfrac}
\be\label{reu:2.21}
P(T) = \int \frac{d^dp}{(2 \pi)^d} \, e^{-p^2 F(p^2) T} \, .
\ee
Introducing $z = p^2$, the final result for the average return probability reads
\be\label{reu:2.22}
P(T) = \frac{1}{(4\pi)^{d/2} \Gamma(d/2)} \int_0^\infty dz \, z^{d/2-1} \, \exp\Big(-z F(z) T \Big) \, , 
\ee
where $F(z) \equiv \bar{\lambda}(k^2 = z)/\bar{\lambda}_{k_0}$. In the classical case, $F(z) = 1$, the relation \eqref{reu:2.22} reproduces the familiar result $P(T) = 1/(4 \pi T)^{d/2}$, whence $\cD_s(T) = d$ independently of $T$. We shall now discuss the spectral dimension for several other illustrative and important examples.
\medskip

\noindent
\textbf{(2) The spectral dimension in QEG.}\index{spectral dimension!in QEG}
\smallskip

\noindent
{\bf (A)} Let us evaluate the average return probability \eqref{reu:2.22} for a simplified RG trajectory where the scale dependence of the cosmological constant is given by a power law, with the same exponent $\delta$ for all values of $k$:
\be\label{reu:2.30}
\bar{\lambda}_k \propto k^\delta \quad \Longrightarrow \quad F(z) \propto z^{\delta/2} \, . 
\ee
By rescaling the integration variable in \eqref{reu:2.22} we see that in this case
\be\label{reu:2.31}
P(T) = \frac{\rm const}{T^{d/(2+\delta)}} \, . 
\ee
Hence \eqref{reu:DsT} yields the important result
\be\label{reu:powerspect}
\framebox{\; \; $\cD_s(T) = \frac{2d}{2+\delta}$ \Big. \; \;} \, .
\ee
It happens to be $T$-independent, so that for $T \rightarrow 0$ trivially
$d_s = 2d/(2+\delta)$.
\smallskip

\noindent
{\bf (B)} Next, let us be slightly more general and assume
that the power law \eqref{reu:2.30} is valid only for squared momenta in a 
certain interval, $p^2 \in [z_1, z_2]$, but $\bar{\lambda}_k$ remains unspecified 
otherwise. In this case we can obtain only partial information about $P(T)$,
namely for $T$ in the interval $[z_2^{-1}, z_1^{-1}]$. The reason is that for $T \in [z_2^{-1}, z_1^{-1}]$
the integral in \eqref{reu:2.22} is dominated by momenta for which 
approximately $1/p^2 \approx T$, i.e., $z \in [z_1, z_2]$. This leads us again to the formula
\eqref{reu:powerspect}, which now, however, is valid only for a restricted range of diffusion times $T$; 
in particular the spectral dimension of interest may not be given by extrapolating \eqref{reu:powerspect} to
$T \rightarrow 0$.
\smallskip

\noindent
{\bf (C)} Let us consider an arbitrary asymptotically safe RG trajectory so that its behavior for $k \rightarrow \infty$ is controlled by
the NGFP. In this case the running of the cosmological constant for $k \gtrsim M$, with $M$ a characteristic mass scale of the order of the Planck mass\index{Planck scale}, is given by a quadratic scale-dependence $\bar{\lambda}_k = \lambda_* k^2$, independently of $d$.
This corresponds to a power law with $\delta = 2$, which entails in the {\bf NGFP regime}, i.e., for $T \lesssim 1/M^2$,
\be\label{reu:UVspec}
\cD_s(T) = \frac{d}{2} \qquad \quad \Big( \mbox{NGFP regime} \Big) \, . 
\ee 
This dimension, again, is locally $T$-independent. It coincides with the $T \rightarrow 0$ limit:
\be
d_s = \frac{d}{2} \, .
\ee
This is the result first derived in ref.\ \cite{reu:oliverfrac}. As it was explained there, it is actually an exact consequence of Asymptotic Safety
which relies solely on the existence of the NGFP and does not depend on the Einstein-Hilbert truncation.
\smallskip

\noindent
{\bf (D)} Returning to the Einstein-Hilbert truncation, let us consider the piece of the Type IIIa RG trajectory depicted in Fig.\ \ref{reu:Fig.epl} which lies inside the linear regime of the GFP. Newton's constant is approximately $k$-independent there and the cosmological constant evolves according to \eqref{reu:linflow}. 
When
$k$ is not too small, so that $\bar{\lambda}_0$ can be neglected relative to $\nu \bar{G} k^d$, we are in what we shall call the ``$k^d$ regime''; it is characterized by a pure power law 
$\bar{\lambda}_k \approx k^\delta$ with $\delta = d$. The physics behind this scale dependence is simple and well-known: It
represents the vacuum energy density obtained by
summing up the zero point energies of all field modes integrated out. For $T$ in the range of scales pertaining to the $k^d$ regime we find
\be
\cD_s(T) = \frac{2d}{2+d} \qquad (k^d \; \mbox{regime}) \, . 
\ee
\smallskip

\noindent
\textbf{(3) The walk dimension in QEG.}\index{walk dimension!in QEG}
In order to determine the walk dimension for the diffusion on the effective QEG space-times
we return to eq.\ \eqref{reu:Heat:FlatSpace} for the untraced heat-kernel. We restrict ourselves to a regime with 
a power law running of $\bar{\lambda}_k$, whence $F(p^2) = (Lp)^\delta$ with some constant length-scale $L$.

Introducing $q_\m \equiv p_\m T^{1/(2+\delta)}$ and $\xi_\m \equiv (x_\m - x_\m^\prime) / T^{1/(2+\delta)}$ we can rewrite \eqref{reu:Heat:FlatSpace}
in the form
\be\label{reu:2.45}
K(x, x^\prime; T) = \frac{1}{T^{d/(2+\delta)}} \, \Phi\left( \frac{|x-x^\prime|}{T^{1/(2+\delta)}} \right)
\ee
with the function
\be
\Phi(|\xi|) \equiv \int \frac{d^dq}{(2\pi)^d} \, e^{i q \cdot \xi} \, e^{- L^\delta q^{2+\delta}} \, .
\ee
For $\delta = 0$, this obviously reproduces \eqref{reu:FSHeat}. From the argument of $\Phi$ in \eqref{reu:2.45} we infer that $r = |x-x^\prime|$ scales as $T^{1/(2+\delta)}$ so that the walk
dimension can be read off as
\be\label{reu:walkspect}
\framebox{\; \; $\cD_w(T) = 2+\delta$ \Big. \; \;} \, . 
\ee
In analogy with the spectral dimension, we use the notation $\cD_w(T)$ rather than $d_w$ to indicate that it might refer to an approximate scaling law which is valid for a finite range of scales only.

For $\delta = 0, 2$, and $d$ we find in particular, for any topological dimension $d$, 
\be\label{reu:2.48}
\cD_w = \left\{ 
\begin{array}{cl}
2   & \mbox{classical regime} \\
4   & \mbox{NGFP regime} \\
2+d \, \,  & k^d\mbox{ regime} \\
\end{array}
\right.
\ee
\index{space-time!classical regime}\index{space-time!semi-classical regime}\index{space-time!NGFP regime}
Regimes with all three walk dimensions of \eqref{reu:2.48} can be realized along a single RG trajectory. Again, the result for the NGFP regime, $\cD_w = 4$, is exact in the sense, that it does not rely on the Einstein-Hilbert truncation.
\medskip

\noindent
\textbf{(4) The Hausdorff dimension in QEG.}\index{Hausdorff dimension!in QEG}
The smooth manifold underlying QEG has per se no fractal properties 
whatsoever. In particular, the volume of a $d$-ball $\cB^d$ covering a patch
of the smooth manifold of QEG space-time scales as
$V(\cB^d) = \int_{\cB^d} d^dx \sqrt{g_k} \propto (r_k)^d \,$. Thus, by comparing
to eq.\ \eqref{reu:Hdd}, we read off that the Hausdorff dimension is strictly equal
to the topological one:
\be\label{reu:hdspect}
\framebox{\; \; $d_H = d$ \Big. \; \;} \, . 
\ee
\smallskip

\noindent
\textbf{(5) The Alexander-Orbach relation.}\index{Alexander-Orbach relation}
For standard fractals the quantities $d_s$, $d_w$, and $d_H$ are not independent but are related by 
\cite{reu:orbach}
\be\label{reu:fracrel}
\frac{d_s}{2} = \frac{d_H}{d_w} \, .
\ee
By combining eqs.\ \eqref{reu:powerspect}, \eqref{reu:walkspect}, and \eqref{reu:hdspect} we see that the same
relation holds true for the effective QEG space-times, at least within the Einstein-Hilbert approximation
and when the underlying RG trajectory is in a regime with power-law scaling of $\bar{\lambda}_k$. For every value
of the exponent $\delta$ we have
\be\label{reu:rel}
\frac{\cD_s(T)}{2} = \frac{d_H}{\cD_w(T)} \, . 
\ee

\noindent
\textbf{(6) (Non-) Recurrence.}
The results $d_H = d$, $\cD_w = 2 + \delta$ imply that, as soon as
$\delta > d-2$, we have $\cD_w > d_H$ and the random walk\index{random walk} is {\it recurrent} then \cite{reu:avra}.
Classically ($\delta = 0$) this condition is met only in low dimensions $d < 2$, but 
in the case of the QEG space-times it is always satisfied in the $k^d$ regime $(\delta = d)$, for example.
So also from this perspective the QEG space-times, due to the specific quantum gravitational dynamics
to which they owe their existence, appear to have a dimensionality smaller than their topological one.
\medskip

\noindent
\textbf{(7) Four dimensions are special.}
It is intriguing that, in the NGFP regime, $\cD_w = 4$ independently of $d$. Hence the walk
is recurrent $(\cD_w > d_H)$ for $d < 4$, non-recurrent for $d > 4$, and the marginal case $\cD_w = d_H$ is realized
if and only if $d=4$, making $d=4$ a distinguished value.

Notably, there is another feature of the QEG space-times which singles out $d=4$: It is the only
dimensionality for which $\cD_s$(NGFP regime)$=d/2$ coincides with the effective dimension
$d_{\rm eff} = d + \eta_* = 2$ obtained from the scale-dependent
graviton propagator (see Sect.\ \ref{reu:sect:7}.)

\section{The RG running of $\cD_s$ and $\cD_w$}
\label{reu:sect:7e}
%
Let us consider an arbitrary RG trajectory $k \mapsto (g_k, \lambda_k)$, where $g_k \equiv G_k k^{d-2}$ and $\lambda_k \equiv \bar{\lambda}_k k^{-2}$ are the dimensionless Newton constant\index{Newton's constant} and cosmological constant
\index{cosmological constant}, respectively. Along such a RG trajectory there might be isolated intervals of $k$-values where the cosmological constant 
evolves according to a power law, $\bar{\lambda}_k \propto k^\delta$, for some constant exponents $\delta$ which are not necessarily the same on different such
intervals. If the intervals are sufficiently long, it is meaningful to ascribe a spectral and walk dimension
\index{walk dimension}\index{walk dimension!running} to them since $\delta = {\rm const}$  implies $k$-independent values 
$\cD_s = 2d/(2+\delta)$ and $\cD_w = 2 + \delta$.

In between the intervals of approximately constant $\cD_s$ and $\cD_w$, where the $k$-dependence of $\bar{\lambda}_k$ is not a
power law,
the notion of a spectral or walk dimension might not be meaningful. The concept of a {\it scale-dependent} dimension
$\cD_s$ or $\cD_w$ is to some extent arbitrary with respect to the way it interpolates between the ``plateaus'' on which $\delta = {\rm const}$  
for some extended period of RG time. While RG methods allow the computation of the $\cD_s$ and $\cD_w$ values on the various plateaus,
it is a matter of convention how to combine them into continuous functions $k \mapsto \cD_s(k), \cD_w(k)$ which interpolate between the respective values.
\smallskip

\noindent
{\bf (1) The exponent $\delta$ as a function on theory space.}
Next we describe a special proposal for a $k$-dependent $\cD_s(k)$ and $\cD_w(k)$ which is motivated by technical simplicity and the general insights
it allows. We retain eqs.\ \eqref{reu:powerspect} and \eqref{reu:walkspect}, but promote $\delta \rightarrow \delta(k)$ to a $k$-dependent quantity 
\be\label{reu:defscale}
\delta(k) \equiv k \p_k \ln(\bar{\lambda}_k) \, .
\ee
When $\bar{\lambda}_k$ satisfies a power law, $\bar{\lambda}_k \propto k^\delta$ this relation reduces to the case of constant
$\delta$.
If not, $\delta$ has its own scale dependence, but no direct physical interpretation should be attributed to it. The particular definition \eqref{reu:defscale} has the special property that it actually can be evaluated without first solving for the RG trajectory. The function
$\delta(k)$ can be seen as arising from a certain scalar function on theory space, $\delta = \delta(g, \lambda)$, whose $k$-dependence results from inserting an RG trajectory: $\delta(k) \equiv \delta(g_k, \lambda_k)$. In fact, \eqref{reu:defscale} implies
$\delta(k) = k \p_k \ln(k^2 \lambda_k) = 2 + \lambda_k^{-1} k \p_k \lambda_k$
so that
$\delta(k) = 2 + \lambda^{-1}_k \beta_\lambda(g_k, \lambda_k)$ upon using the RG-equation $k \p_k \lambda_k = \beta_\lambda(g, \lambda)$. Thus when we consider
$\delta$ as a function on theory space, coordinatized by $g$ and $\lambda$, it reads
\be\label{reu:deltatheoryspace}
\delta(g, \lambda) = 2 + \frac{1}{\lambda} \, \beta_\lambda(g, \lambda) \, . 
\ee
Substituting this relation into
 \eqref{reu:powerspect} and \eqref{reu:walkspect}, the spectral and the walk dimensions become functions
on the $g$-$\lambda$-plane
\be\label{reu:dstheo}
\cD_s(g, \lambda) = \frac{2d}{4 + \lambda^{-1} \beta_\lambda(g, \lambda)} \, , 
\ee
and
\be\label{reu:dwtheo}
\cD_w(g, \lambda) = 4 + \lambda^{-1} \beta_\lambda(g, \lambda) \, .
\ee

\begin{figure}[t]
	\centering
	\includegraphics[width=0.78\textwidth]{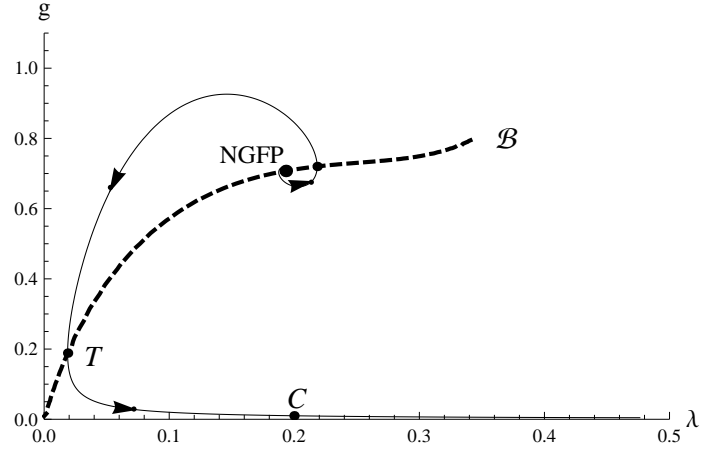}
	\caption{The $g$-$\lambda-$theory space with the line of turning points, $\cB$, and a
		typical trajectory of Type IIIa. The arrows point in the direction of decreasing $k$. The big
		black dot indicates the NGFP while the smaller dots represent points at which the RG trajectory switches
		from increasing to decreasing $\lambda$ or vice versa. The point $T$ is the lowest turning point,
		and $C$ is a typical point within the classical regime. For $\lambda \gtrsim 0.4$, the RG flow leaves
		the classical regime and is no longer reliably captured by the Einstein-Hilbert truncation.}
  \label{reu:Fig.epl}
\end{figure}
As we discussed already, the scaling regime of a NGFP has the exponent $\delta = 2$. From eq.\ \eqref{reu:deltatheoryspace} we learn that this value is realized at all points $(g, \lambda)$ where $\beta_\lambda =0$. The second condition for the NGFP, $\beta_g = 0$, is not required here, so that we have $\delta = 2$ along the entire line in theory space:
\be\label{reu:2.75}
\cB = \Big\{ \, (g, \lambda) \, \Big| \, \beta_\lambda(g, \lambda) = 0 \, \Big\} \, . 
\ee
For $d=4$ the curve $\cB$ is shown as the dashed line in Fig.\ \ref{reu:Fig.epl}. Both the GFP $(g, \lambda) = (0, 0)$ and the NGFP, $(g, \lambda) = (g^*, \lambda^*)$, are located on this curve. Furthermore, the turning points
$T$ of all Type IIIa trajectories are also situated on $\cB$, and the same holds for all the higher order turning points which occur when the trajectory spirals around the NGFP. The line $\cB$ divides the $g$-$\lambda$--plane in three domains: (i) Above
$\cB$: $\beta_\lambda>0$, $\delta>2 \; \Rightarrow \; \cD_s<d/2,\;\cD_w>4$. (ii) Below $\cB$: $\beta_\lambda<0$, $\delta<2 \;
\Rightarrow \; \cD_s>d/2,\;\cD_w<4$. (iii) On $\cB$: $\beta_\lambda=0$, $\delta=2 \; \Rightarrow \; \cD_s=d/2,\;\cD_w=4$.
This observation leads us to an important conclusion: The values $\delta = 2 \Longleftrightarrow  \cD_s = d/2, \cD_w = 4$ which (without involving any truncation) are found in the NGFP regime, actually also apply to all points $(g, \lambda) \in \cB$, provided the Einstein-Hilbert truncation is reliable and no matter is included.
\smallskip

\noindent
{\bf (2) Running dimensions along a RG trajectory.}
We proceed by investigating how the spectral and walk dimension of the effective QEG space-times changes along a given RG trajectory.
As discussed above, our interest is in scaling regimes where $\cD_s$ and $\cD_w$ remain (approximately) constant for a long interval of $k$-values.
For the remainder of this section, we will restrict ourselves to the Einstein-Hilbert truncation in $d=4$. 

We start by numerically solving the coupled differential equations \eqref{reu:EHflow}
with the $\beta$-functions from \cite{reu:mr} for a series of initial conditions keeping
$\lambda_{\rm init} = \lambda(k_0) = 0.2$ fixed and successively lowering
$g_{\rm init} = g(k_0)$. The result is a family of RG trajectories where the classical regime becomes more and more pronounced. Subsequently, these solutions are substituted into \eqref{reu:dstheo} and \eqref{reu:dwtheo}, which give $\cD_s(t; g_{\rm init}, \lambda_{\rm init})$ and $\cD_w(t; g_{\rm init}, \lambda_{\rm init})$ in dependence of the RG-time $t \equiv \ln(k)$ and the RG trajectory. One can verify explicitly, that substituting the RG trajectory into the return probability \eqref{reu:2.22} and computing the spectral dimension\index{spectral dimension}\index{spectral dimension!running} from \eqref{reu:DsTlim} by carrying out the resulting integrals numerically gives rise to the same picture.

\begin{figure}[t]
	\begin{minipage}{.49\columnwidth}
		\centering
		\includegraphics[width=0.99\columnwidth]{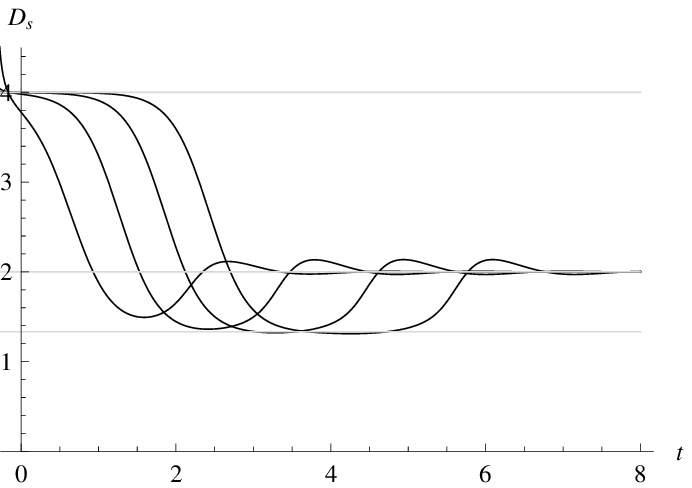}
	\end{minipage}
	\hfill
	\begin{minipage}{.49\columnwidth}
		\centering
		\includegraphics[width=0.99\columnwidth]{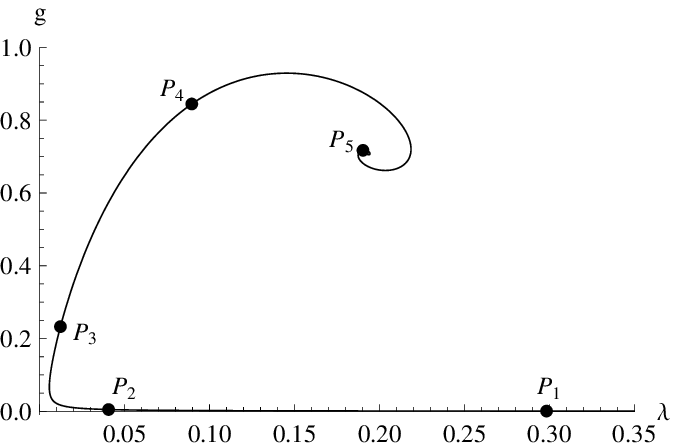}
	\end{minipage}
\caption{The $t \equiv \ln(k)$-dependent spectral dimension along illustrative solutions of the RG-equations
	\eqref{reu:EHflow} in $d=4$. The trajectories develop three plateaus: the classical plateau with $\cD_s =4, \cD_w = 2$,
	the semi-classical plateau where $\cD_s = 4/3, \cD_w = 6$ and the NGFP plateau with $\cD_s = 2, \cD_w = 4$.
	\index{space-time!classical regime}\index{space-time!semi-classical regime}\index{space-time!NGFP regime}
	(Recall that $\cD_w=2d/\cD_s=8/\cD_s$.) The plateau	values are indicated by the horizontal lines. The second figure shows
	the location of these plateaus on the RG trajectory: the classical, $k^4$, and NGFP regime appear between the points $P_1$
	and $P_2$, $P_3$ and $P_4$, and above $P_5$, respectively.}
\label{reu:Fig.spec}
\end{figure}

Fig.\ \ref{reu:Fig.spec} shows the resulting spectral dimension and the localization of the plateau-regimes on the RG
trajectory. In the left diagram, $g_{\rm init}$ decreases by one order of magnitude for each shown trajectory, starting with the highest value to the very left.
As a central result, Fig.\ \ref{reu:Fig.spec} establishes that the RG flow gives rise to {\it three} plateaus where $\cD_s(t)$ and $\cD_w(t)$ are approximately constant: \\
{\bf (i)} For small values $k$, below $t \simeq 1.8$, say, one finds a {\it classical plateau} where $\cD_s = 4, \cD_w = 2$ for a long range of $k$-values. Here $\delta = 0$, indicating that the cosmological constant is indeed constant. \\
{\bf (ii)} Following the RG flow towards the UV (larger values of $t$) one next encounters the {\it semi-classical plateau} where $\cD_s = 4/3, \cD_w = 6$. In this case $\delta(k) = 4$ so that $\bar{\lambda}_k \propto k^4$ on the corresponding part of the RG trajectory. \\
{\bf (iii)} Finally, the {\it NGFP plateau} is characterized by $\cD_s = 2, \cD_w = 4$, which results from the scale-dependence of the cosmological constant at the NGFP $\bar{\lambda}_k \propto k^2 \Longleftrightarrow  \delta = 2$.

The plateaus become more and more extended the closer the trajectory's turning point $T$ gets to the GFP, i.e., the smaller
the IR value of the cosmological constant.

\section{Matching the spectral dimensions of QEG and CDT}
\label{reu:sect:7f}
The key advantage of the spectral dimension $\cD_s(T)$ is that it may 
be defined and computed within various a priori unrelated approaches to
quantum gravity. In particular, it is easily accessible in Monte Carlo simulations
of the Causal Dynamical Triangulations (CDT)\index{CDT}\index{Causal Dynamical Triangulations|see{CDT}}
\index{CDT!comparison with QEG} approach in $d=4$ \cite{reu:ajl34} and $d=3$ \cite{reu:Benedetti:2009ge} as well as in
Euclidean Dynamical Triangulations (EDT) \cite{reu:laiho-coumbe}. This feature allows a direct comparison between
$\cD_s^{\rm CDT}(T)$ and $\cD_s^{\rm EDT}(T)$ obtained within the discrete approaches and $\cD_s^{\rm QEG}(T)$ capturing the
fractal properties of the QEG effective space-times. In \cite{reu:frankfrac} we carried out this analysis for $d=3$, using
the Monte Carlo data obtained in \cite{reu:Benedetti:2009ge} according to the following scheme: 

\noindent
{\bf (i)} First, we numerically construct a RG trajectory $g_k(g_0, \lambda_0), \lambda_k(g_0, \lambda_0)$ depending on the initial conditions $g_0, \lambda_0$, by solving the flow equations \eqref{reu:EHflow}. \\
{\bf (ii)} We evaluate the resulting spectral dimension $\cD_s^{\rm QEG}(T; g_0, \lambda_0)$ of the corresponding effective
QEG space-time. This is done by first finding the return probability $P(T; g_0, \lambda_0)$, eq.\ \eqref{reu:2.22}, for the
RG trajectory under consideration and then substituting the resulting expression into \eqref{reu:DsT}.  Besides on the length
of the random walk, the spectral dimension constructed in this way also depends on the initial conditions of the RG
trajectory. \\
{\bf (iii)} We determine the RG trajectory underlying the CDT-simulations by 
fitting the parameters $g_0, \lambda_0$ to the Monte Carlo data. The corresponding best-fit values are obtained via an ordinary least-squares fit, minimizing the squared Euclidean distance 
\be\label{reu:lsf}
(\Delta \cD_s )^2 \equiv \sum_{T = 20}^{500} \, \left( \cD_s^{\rm QEG}(T; g_0^{\rm fit}, \lambda_0^{\rm fit}) - \cD_s^{\rm CDT}(T) \right)^2 \, , 
\ee
 between the (continuous) function $\cD_s^{\rm QEG}(T; g_0, \lambda_0)$ and the points $\cD_s^{\rm CDT}(T)$. We thereby restrict ourselves to 
the random walks with discrete, integer length $20 \le T \le 500$, which constitute the ``reliable'' part of the data. 

\begin{table}
\caption{Initial conditions $g_0^{\rm fit}, \lambda_0^{\rm fit}$ for the RG trajectory providing
	the best fit to the Monte Carlo data \cite{reu:Benedetti:2009ge}. The fit-quality $(\Delta \cD_s )^2$,
	given by the sum of the squared residues, improves systematically when increasing the number of simplices
	in the triangulation.}
\label{reu:Table.1}
\renewcommand{\arraystretch}{1.5}
\begin{center}
\begin{tabular}{c c c c }
\hline
        & \quad \qquad $g_0^{\rm fit}$ \qquad \quad & \quad \qquad $\lambda_0^{\rm fit}$ \qquad \qquad & \quad \quad  \quad $(\Delta \cD_s )^2$ \quad  \quad \quad \\ \svhline
 \quad $70$k\quad &  $0.7 \times 10^{-5}$ & $ 7.5 \times 10^{-5}$  & $0.680$ \\
 \quad$100$k\quad	&  $8.8 \times 10^{-5}$ & $39.5 \times 10^{-5}$  & $0.318$ \\
 \quad$200$k\quad &  $13 \times 10^{-5}$  & $61 \times 10^{-5}$    & $0.257$ \\ \hline
\end{tabular}
\end{center}
\renewcommand{\arraystretch}{1}
\end{table}
The resulting best-fit values $g_0^{\rm fit}, \lambda_0^{\rm fit}$ for triangulations with $N = 70.000$, $N=100.000$, and $N=200.000$ simplices are collected in Table \ref{reu:Table.1}.
Notably, the sum over the squared residuals in the third column of the table  
 improves systematically with an increasing number of simplices. By integrating
 the flow equation for $g(k), \lambda(k)$ for the best-fit initial conditions one furthermore observes that the points $g_0^{\rm fit}, \lambda_0^{\rm fit}$ are actually located 
on {\it different} RG trajectories. Increasing the size of the simulation $N$ leads to a mild, but systematic increase of the distance between the turning point $T$ and the GFP of the corresponding best-fit trajectories.

\begin{figure}[t]
	\centering
	\includegraphics[width=0.47\textwidth]{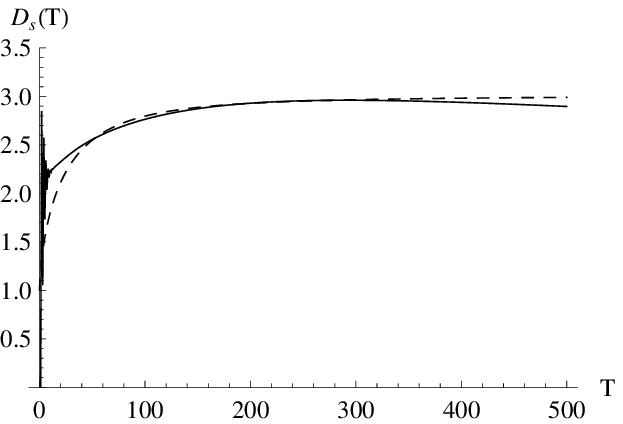} \;\,
	\includegraphics[width=0.47\textwidth]{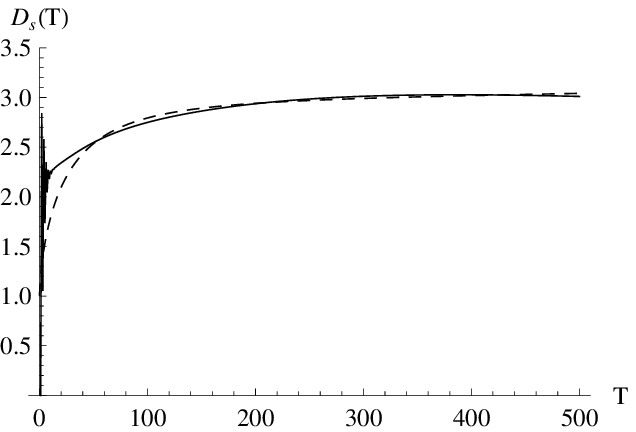} \\
	\includegraphics[width=0.47\textwidth]{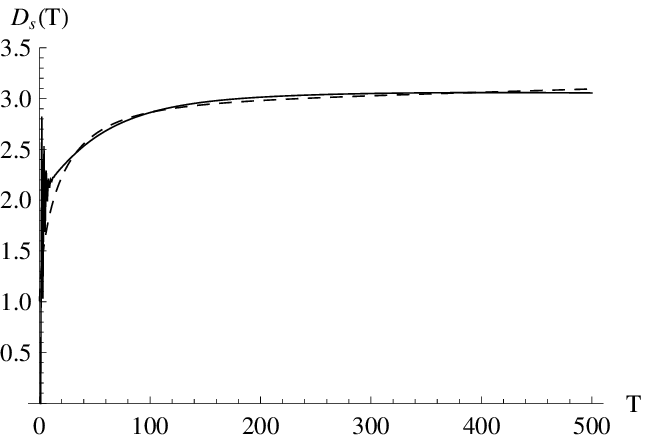} \;\,
	\includegraphics[width=0.47\textwidth]{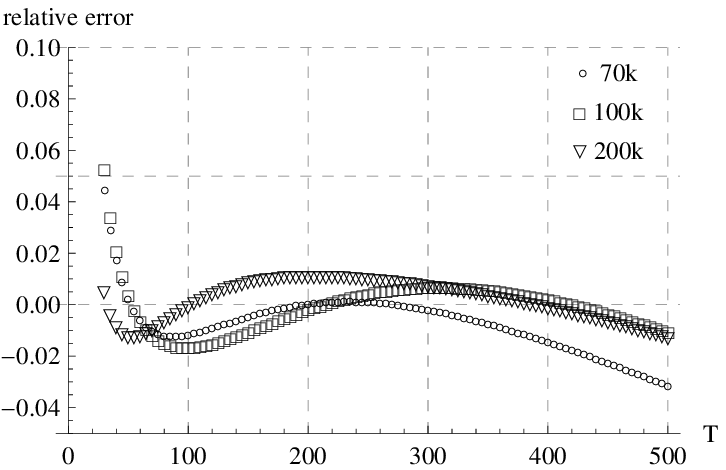}
\caption{Comparison between the spectral dimension measured in 3-dimensional CDT space-times build from $70$k (upper left), $100$k (upper right),
	and $200$k simplices (lower left) obtained in \cite{reu:Benedetti:2009ge} (continuous curves) and the best fit values for
	$\cD_s^{\rm QEG}(T; g_0^{\rm fit}, \lambda_0^{\rm fit})$ (dashed curves). The relative errors for the fits
	to the CDT-datasets with $N = 70.000$ (circles), $N=100.000$ (squares) and $N = 200.000$ (triangles) simplices
	are shown in the lower right. The residuals grow for very small and very large durations $T$ of the random walk,
	consistent with discreteness effects at small distances and the compactness of the simulation for large values
	of $T$, respectively. The quality of the fit improves systematically for triangulations containing more
	simplices. For the $N=200$k data the relative error is $\approx 1\%$.}
\label{reu:p.fit1}
\end{figure}

Fig.\ \ref{reu:p.fit1} then shows the direct comparison between the spectral dimensions obtained by the simulations (continuous curves) and the best-fit QEG trajectories (dashed curves)
for $70$k, $100$k and $200$k in the upper left, upper right and lower left panel, respectively. 
 This data is complemented by the relative error 
\be
\epsilon \equiv - \frac{\cD_s^{\rm QEG}(T; g_0^{\rm fit}, \lambda_0^{\rm fit}) - \cD_s^{\rm CDT}(T)}{\cD_s^{\rm QEG}(T; g_0^{\rm fit}, \lambda_0^{\rm fit})}
\ee
for the three fits in the lower right panel. The $70$k data still shows a systematic deviation from the classical value $\cD_s(T) = 3$ for long random walks, which is not present in the QEG results. This mismatch decreases systematically for larger triangulations where the classical regime becomes more and more pronounced. Nevertheless and most remarkably we find that for the $200$k-triangulation  $\epsilon \lesssim 1\%$, throughout.

\begin{figure}[t]
	\centering
	\includegraphics[width=.9\columnwidth]{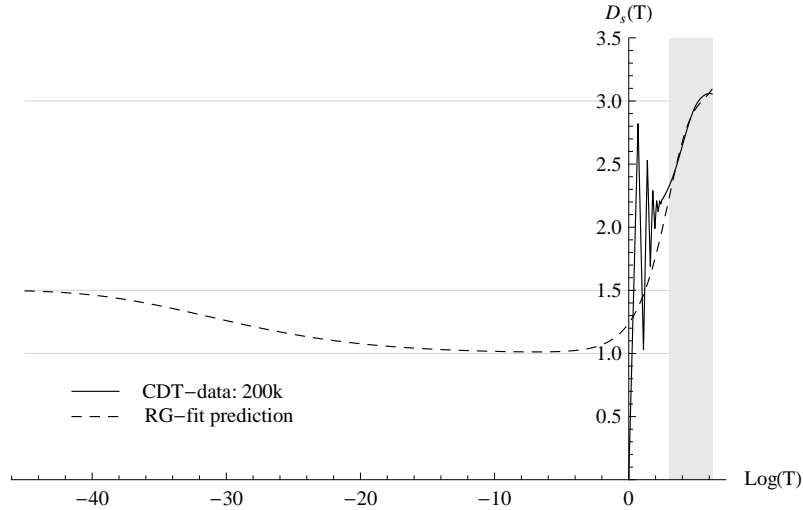}
\caption{Comparison between the spectral dimensions obtained from the dynamical triangulation with
	$200$k simplices (continuous curve) and the corresponding $\cD_s^{\rm QEG}(T; g_0^{\rm fit}, \lambda_0^{\rm fit})$
	predicted by QEG (dashed curve). In the latter case, the scaling regime corresponding to the NGFP is reached
	for $\log(T) < -40$, which is well below the distance scales probed by the Monte Carlo simulation.}
\label{reu:p.fit2}
\end{figure}

We conclude this section by extending $\cD_s^{\rm QEG}(T; g_0^{\rm fit}, \lambda_0^{\rm fit})$ obtained from the $200$k data to the region of very short random walks $T < 20$. The result is depicted in Fig.\ \ref{reu:p.fit2} which displays $\cD_s^{\rm CDT}(T)$ (continuous curve) and $\cD_s^{\rm QEG}(T; g_0^{\rm fit}, \lambda_0^{\rm fit})$ (dashed curve) as a function of
$\log(T)$.\index{CDT} 
Similarly to the four-dimensional case discussed in Fig.\ \ref{reu:Fig.spec}, the function $\cD_s^{\rm QEG}(T; g_0^{\rm fit}, \lambda_0^{\rm fit})$ obtained for $d=3$ develops three plateaus where the spectral dimension is approximately constant over a long $T$-interval. For successively decreasing duration of the random walks, these plateaus correspond to the classical regime $\cD_s^{\rm QEG}(T) = 3$, the semi-classical regime where $\cD_s^{\rm QEG}(T) \approx 1$ and the NGFP regime where $\cD_s^{\rm QEG}(T) = 3/2$. The figure illustrates that $\cD_s^{\rm CDT}(T)$ probes the classical regime and part of the first crossover towards the semi-classical regime only. This is in perfect agreement with the assertion \cite{reu:Benedetti:2009ge} that the present simulations do not yet probe structures below the Planck scale. This assessment resolves the apparent contradiction between the extrapolation result $\lim_{T \rightarrow 0} \cD_s^{\rm CDT}(T) \approx 2$ and the QEG prediction $\lim_{T \rightarrow 0} \cD_s^{\rm QEG}(T) = 3/2$. Performing the extrapolation of $\lim_{T \rightarrow 0} \cD_s^{\rm CDT}(T)$ based on the leading corrections to the classical regime does not reliably identify the signature of a non-Gaussian fixed point in $\cD_s(T)$.

A similar conclusion also holds true in four dimensions. Comparing the profiles of $\cD_s^{\rm QEG}(T)$ shown in Fig.\ \ref{reu:Fig.spec} with the fitting functions used in the CDT \cite{reu:ajl34} or EDT \cite{reu:laiho-coumbe} simulations shows that all the Monte Carlo data points obtained are positioned on the {\it infrared} side of the turning point of the RG trajectories underlying the QEG effective space-times. They neither probe the semi-classical plateau nor the scaling regime of the NGFP. Depending on where the data is cut off, one obtains different tangents to the first crossover, which lead to widely different extrapolations for the value $d_s = \cD_s(T)|_{T=0}$. We believe that this is actually at the heart of the apparent mismatch in the spectral dimension for infinitesimal random walks reported from the CDT and EDT computations.

\section{Asymptotic Safety in cosmology}
\label{reu:cosmo}
At this point we switch to another field where QEG effects might be relevant, the cosmology of the early Universe.
As we discussed at the end of Sect.\ \ref{reu:sect:4}, the Type IIIa trajectories displayed in Fig.\ \ref{reu:fig0} 
 possess all the qualitative
properties one would expect from the  RG trajectory describing gravitational phenomena
in the real Universe\index{early Universe}. They can have a long classical regime and a small,
positive cosmological constant\index{cosmological constant} in the infrared (IR).
Remarkably, along the RG trajectory realized by Nature \cite{reu:h3,reu:entropy},
the dimensionful running cosmological
constant $\lbar(k)$ changes by about 120 orders of magnitude between $k$-values of the order
of the Planck mass and macroscopic scales, while the dimensionful Newton constant $G(k)$\index{Newton's constant} has no
strong $k$-dependence in this regime. For $k> \mp$, the scale-dependence of $G(k)$ and $\lbar(k)$ is governed by the NGFP,
implying that $\lbar(k)$ diverges and $G(k)$ approaches zero, see eq.\ \eqref{reu:asymrun}.
An immediate question is whether there is any experimental or observational 
evidence that would  hint at this enormous scale dependence of the gravitational parameters. 
Clearly the natural place to search for such phenomena
is cosmology.
\medskip

\noindent
\textbf{(1) RG improved Einstein equations.}\index{improved Einstein equations}
The computational setting for investigating the signatures arising from the scale-dependent couplings
are the RG improved Einstein equations: By means of a suitable cutoff
identification $k=k(t)$ we turn the scale dependence of $G(k)$ and $\lbar(k)$
into a time dependence, and then substitute the resulting $G(t)\equiv G(k(t))$
and $\lbar(t)\equiv \lbar(k(t))$ into the Einstein equations
$G_{\mu\nu}=-\lbar(t)g_{\mu\nu}+8\pi G(t) T_{\mu\nu}$. 
We specialize $g_{\mu\nu}$
to describe a spatially flat $(K=0)$ Robertson-Walker\index{Robertson-Walker geometry} metric with scale factor $a(t)$,
and we take ${T_\mu}^\nu = {\rm diag}[-\rho,p,p,p]$ to be the energy momentum
tensor of an ideal fluid with equation of state $p=w\rho$ where $w>-1$ is constant.
Then the improved Einstein equation boils down 
to the modified Friedmann equation\index{modified Friedmann equation} and a continuity equation:
\begin{subequations}
\ba\label{reu:2.5a}
&&H^2 = \frac{8\pi}{3} G(t) \; \rho + \frac{1}{3}\lbar(t)\,,\\[2mm]
&&\dot\rho+3H(\rho+p)=-\frac{\dot{\lbar}+8\pi \; \rho \; \dot{G}}{8\pi \; G}\;.\label{reu:2.5b}
\ea
\end{subequations}
The modified continuity equation (\ref{reu:2.5b}) is the integrability condition for the 
improved Einstein equation implied by Bianchi's identity, 
$D^\mu [\lbar(t) g_{\mu\nu} - 8\pi G(t) T_{\mu\nu}] =0$. 
It describes the energy exchange 
between the matter and gravitational degrees of freedom (geometry).
For later use let us note that upon
defining the critical density
$\rho_{\rm crit}(t)\equiv {3 \; H(t)^2}/{8\pi \; G(t)}$,
the relative density $\Omega_{\rm M}\equiv \rho/\rho_{\rm crit}$ and 
$\Omega_{\lbar}=\rho_{\lbar}/\rho_{\rm crit}$ the modified Friedmann equation (\ref{reu:2.5a})
can be written as
$\OM(t) +\OL(t) = 1$.
\medskip

\noindent
\textbf{(2) Solving the RG improved Einstein Equations.}\index{improved Einstein equations}
The general strategy for solving eqs.\ (\ref{reu:2.5a}, \ref{reu:2.5b})
is as follows. First we obtain $G(k)$ and $\lbar(k)$ by solving the flow equation in the Einstein-Hilbert truncation before
 constructing the cosmologies by numerically solving the RG improved evolution equations.
We shall employ the cutoff identification $k(t) =\xi H (t)$,
where $\xi$ is a fixed positive constant of order unity.
This is a natural choice since in a Robertson-Walker geometry
the Hubble parameter\index{Hubble parameter} measures the curvature of space-time; its inverse $H^{-1}$ defines 
the size of the ``Einstein elevator''. 

The very early part of the cosmology can be described analytically. For $k\rightarrow \infty$ the trajectory
approaches the NGFP\index{NGFP} so that $G(k)=g^\ast/k^2$ and 
$\lbar(k)=\lambda^\ast k^2$. In this case the differential equation can be solved analytically, with the result
\be\label{reu:NGFPsolution}
H(t)=\alpha/t, \;\;\;\; a(t) = At^\alpha, \;\;\;\;\; \alpha = \Big [ \frac{1}{2}(3+3w)(1-\OLS) \Big]^{-1}\,,
\ee
and $\rho(t)=\widehat\rho t^{-4}, \;\; G(t) = \widehat G  t^2,  \;\;  \lbar(t) = \widehat{\lbar}  /t^2$.
Here $A$, $\widehat\rho$, $\widehat G $, and $\widehat{\lbar} $ are positive constants.
They parametrically depend on the relative
vacuum energy density in the fixed point regime, $\OLS$, which assumes
values in the interval $(0,1)$.
If $\alpha>1$ the deceleration parameter $q=\alpha^{-1} -1$ is  negative and the 
Universe is in a phase of {\it power law inflation}. Furthermore, it has 
{\it no particle horizon} if
 $\alpha \geq 1$, but does have a horizon of radius $d_{H}=t/(1-\alpha)$
if $\alpha < 1$. In the case of $w=1/3$ this means that there is  a horizon for 
$\OLS< 1/2$, but none if $\OLS \geq 1/2$.
\medskip

\noindent
\textbf{(3) Inflation in the fixed point regime.}
Next we discuss in more detail the cosmologies originating from the epoch of power law inflation which is realized
in the NGFP regime if $\OLS > 1/2$. Since 
the transition from the fixed point to the classical FRW regime is rather sharp, it will be
sufficient to approximate the RG improved UV cosmologies by the following caricature :
For $0<t< t_{\rm tr}$, the scale factor behaves as
$a(t)\propto t^\alpha$, $\alpha > 1$.  Here $\alpha = (2-2\OLS)^{-1}$ since $w =1/3$ 
will be assumed. Thereafter, for $t>t_{\rm tr}$, we have a classical, entirely
matter-driven expansion $a(t)\propto t^{1/2}$ . Clearly this is a very attractive scenario:
{\it neither to trigger inflation nor to stop it one needs any ad hoc ingredients such
as an inflaton\index{inflaton} field or a special potential}. It suffices to include the leading quantum
effects in the gravity + matter system. Following \cite{reu:entropy}, the RG improved cosmological
evolution for the RG trajectory realized by Nature is characterized as follows:
\medskip

\noindent
\textbf{(A)}
The transition time $t_{\rm tr}$ is dictated by the RG trajectory. 
It leaves the asymptotic scaling 
regime near $k\approx \mp$. Hence  $H(t_{\rm tr})\approx \mp$ and since $\xi=O(1)$ and $H(t)=\alpha/t$,
we find the estimate
\be\label{reu:6.2}
t_{\rm tr}= \alpha \; \tp\,.
\ee
Here, as always, the Planck mass, time, and length\index{Planck scale} are defined in terms of the value of Newton's
constant in the classical regime :
$\tp = \lp = \mp^{-1} = \bar{G}^{1/2} = G_{\rm observed}^{1/2}$.
Let us now assume that $\OLS$ is very close 
to $1$ so that $\alpha$ is large:
$\alpha \gg 1$. Then (\ref{reu:6.2}) implies that the transition takes place at a cosmological time\index{cosmological time}
which is much later than the Planck time. At the transition the {\it Hubble parameter} is of order $\mp$, but the
{\it cosmological time} is in general not of the order of $\tp$. Stated differently, the ``Planck time''
is {\it not} the time at which $H$ and the related physical quantities assume Planckian values. 
The Planck time as defined above is well within the NGFP regime:
$\tp = t_{\rm tr} / \alpha \ll t_{\rm tr}$. 

\begin{figure}[t]
\begin{center}
\includegraphics[width=.8\columnwidth]{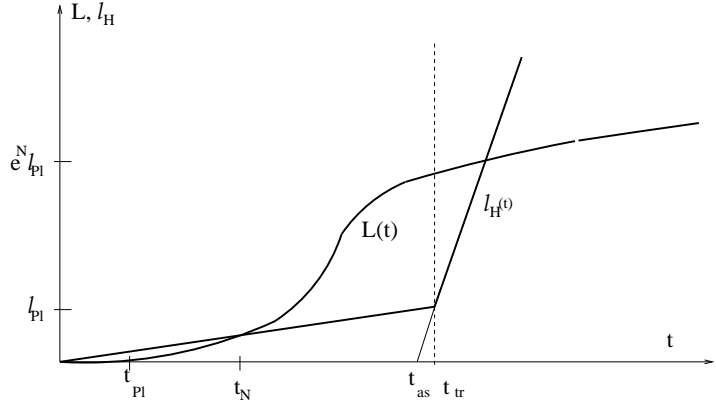}
\end{center}
\caption{Shown is the proper length $L$ and the Hubble radius as a function of time. 
The NGFP and FRW cosmologies are valid for $t<t_{\rm tr}$ and $t>t_{\rm tr}$, respectively.
The classical cosmology has an apparent initial singularity at $t_{as}$ outside its domain of 
validity. Structures of size $e^N \lp$ at $t_{\rm tr}$ cross the Hubble radius at $t_N$, a 
time which can be larger than the Planck time. }
\label{reu:fig9}
\end{figure}

In the NGFP regime $0<t< t_{\rm tr}$ the Hubble radius $\ell_{H} (t) \equiv 1/H(t)$, i.e.,
$\ell_{H} (t) = t/{\alpha} $, 
increases linearly with time but, for $\alpha \gg 1$, with a very small slope. At the transition, $t=t_{\rm tr}$, 
the NGFP solution is to be matched continuously with a FRW cosmology
(with vanishing cosmological constant). We may use the  classical formula $a\propto \sqrt{t}$ 
for the scale factor, but we must shift the time axis on the classical side such that $a$, 
$H$, and then as a result of (\ref{reu:2.5a}), also $\rho$ are continuous at $t_{\rm tr}$. Therefore,
$a(t)\propto (t-t_{\rm as})^{1/2}$ and 
$H(t) = \frac{1}{2} \; (t-t_{\rm as})^{-1} \;\;\; \text{for } \;\;\; t> t_{\rm tr}$.
Equating this Hubble parameter  at $t=t_{\rm tr}$ to 
$H(t) = \alpha /t$, valid in the NGFP regime, we find that the shift $t_{\rm as}$ must be chosen as
$t_{\rm as} =  (\alpha -\frac{1}{2}) \tp = (1 - \frac{1}{2\alpha})  t_{\rm tr}  <  t_{\rm tr}$.
Here the subscript 'as' stands for ``apparent singularity''. This is to indicate that if one continues
the classical cosmology to times $t<t_{\rm tr}$, it has an initial singularity (``big bang'') at 
$t=t_{\rm as}$. Since, however, the FRW solution is not valid there, nothing special happens at 
$t_{\rm as}$; the true initial singularity is located at $t=0$ in the NGFP regime, see Fig.~\ref{reu:fig9}.
\medskip

\noindent
\textbf{(B)}
We now consider some structure of comoving length $\Delta x$, 
a single wavelength of a density perturbation,
for instance. The corresponding physical, i.e., proper length is $L(t) = a(t) \Delta x$ then.
In the NGFP regime it has the time dependence $L(t) =  ({t}/{\tr}  )^{\alpha} \; L(\tr)$.
The ratio of $L(t)$ and the Hubble radius evolves according to 
$L(t)/\LH (t) =  ( t/\tr  )^{\alpha-1} \, L(\tr)/\LH (\tr)$.
For $\alpha > 1$, i.e., $\OLS > 1/2$, the proper length of any object grows faster than the Hubble
radius. So objects which are of ``sub-Hubble'' size at early times can cross the Hubble radius and become
``super-Hubble'' at later times, see Fig.~\ref{reu:fig9}. 

Let us focus on a structure which, at $t=\tr$, is
$e^N$ times larger than the Hubble radius. Before the transition we have
$L(t)/\LH (t) = e^N \; (t/\tr)^{\alpha -1}$.
Assuming $e^N > 1$, there exists a time $t_N < \tr$ at which $L(t_{N}) =\LH (t_{ N})$
so that the structure considered ``crosses'' the Hubble radius at the time $t_N$. It is given
by 
\be\label{reu:6.8}
t_{N}=\tr \;{\rm exp} {\Big ( -\frac{N}{\alpha -1} \Big )}\,.
\ee
What is remarkable about this result is that, even with rather moderate values of $\alpha$, one can 
easily ``inflate'' structures to a size which is by many $e$-folds larger than the Hubble radius 
{\it during a very short time interval at the end of the NGFP epoch}. 

The largest structures in the present Universe, evolved backward in time by the classical equations
to the point where $H=\mp$, have a size of about $e^{60}\; \lp$ there. We can use 
(\ref{reu:6.8}) with $N=60$ to find the time $t_{\rm 60}$ at which those structures crossed
the Hubble radius. With $\alpha = 25$ the result is $t_{\rm 60}=2.05\; \tp = \tr /12.2$. Remarkably, $t_{\rm 60}$ is smaller than 
$\tr$ by one order of magnitude only. As a consequence, the physical conditions prevailing at the
time of the crossing are not overly  ``exotic'' yet. 
The Hubble parameter, for instance, is only one order of magnitude larger than at the transition:
$H(t_{\rm 60})\approx 12 \mp$.  The same is true for the temperature; one can show that 
$T(t_{\rm 60})\approx 12 T(\tr)$ where $T(\tr)$ is of the order of $\mp$. Note  that $t_{\rm 60}$ is larger than $\tp$.
\medskip

\noindent
\textbf{(C)}
QEG offers a natural mechanism for generating primordial fluctuations during the NGFP epoch. The idea is that the
NGFP amounts to a kind of ``critical phenomenon'' with characteristic fluctuations on all scales. They turn out
to have a scale free spectrum with a spectral index close to $n=1$.
For a detailed discussion of this mechanism the reader is referred to \cite{reu:oliver2,reu:cosmo1,reu:entropy}.
Suffice it to say that the quantum mechanical generation of the primordial fluctuations makes essential use of the
dimensionally reduced form of the graviton propagator; it
happens on sub-Hubble distance scales. However, thanks to the inflationary NGFP era\index{inflationary NGFP era} the modes
relevant to cosmological structure formation were indeed smaller than the Hubble radius at a sufficiently early time,
for $t<t_{\rm 60}$, say. (See the $L(t)$ curve in Fig.~\ref{reu:fig9}.) 
\medskip

\noindent
\textbf{(4) Entropy and the renormalization group.}\index{entropy|textbf}
In standard Friedmann-Robertson-Walker cosmology where the expansion is adiabatic,
the entropy (within a comoving volume) is constant. It has always been somewhat puzzling therefore
where the huge amount of entropy contained in the present Universe comes from. 
Presumably it is dominated by the CMBR photons which  contribute an amount
of about $10^{88}$ to the entropy within the present Hubble sphere.

The observation that the value of the cosmological constant decreases during
the expansion of the universe hints at another mechanism at work within the 
RG improved cosmologies: the dynamical creation of entropy.
Following \cite{reu:entropy}  we shall argue that in principle the entire entropy of the massless fields in the present
Universe can be understood as arising from this effect.  
If energy can be exchanged freely between
the cosmological constant and the matter degrees of freedom, the entropy observed today 
is obtained precisely if the initial entropy\index{entropy} at the ``big bang'' vanishes.
The assumption that the matter system must allow for an unhindered energy exchange with
$\lbar$ is essential, see  refs. \cite{reu:cosmo1,reu:entropy}.

To make the argument as transparent as possible, let us first consider a Universe without matter,
but with a positive $\lbar$. Assuming maximal symmetry, this is nothing but  de Sitter
space, of course. In static coordinates its metric is given by
$ds^2 = -(1+2\Phi_{\rm N}(r) ) dt^2+ (1+2\Phi_{\rm N}(r))^{-1}dr^2 +
r^2 (d\theta^2 +\sin^2 \theta d\phi^2)$ with $\Phi_{\rm N}(r) = -\frac{1}{6}\, \lbar\, r^2$.
In the weak field and slow motion limit $\Phi_{\rm N}(r)$ has the interpretation
of a Newtonian potential; for $\lbar >0$ it is an upside-down parabola. Point particles in
this space-time ``roll down the hill'' and are rapidly driven away from the origin
$r=0$ and from any other particle. Now assume that the magnitude of $|\lbar|$
is slowly (``adiabatically'') decreased. This will cause the potential $\Phi_{\rm N}(r)$ 
to move upward as a whole, its slope decreases. So the change in $\lbar$ increases the particle's
potential energy.  This is the simplest way of understanding that a {\it positive decreasing}
cosmological constant has the effect of ``pumping'' energy into the matter degrees of freedom.

We are thus led to  suspect that, because of the decreasing cosmological constant, 
there is a continuous inflow of energy into the cosmological fluid contained in an 
expanding Universe. It will ``heat up'' the fluid or, more exactly, lead to a slower decrease
of the temperature than in standard cosmology. Furthermore, by elementary thermodynamics, it will 
{\it increase} the entropy of the fluid. If during the time $dt$ an amount of heat
$d Q>0$ is transferred into a volume $V$ at the temperature $T$ the entropy changes by an amount $dS=dQ/T>0$.
To be as conservative (i.e., close to standard cosmology) as possible, we assume that this process
is reversible. If not, $dS$ is even larger.

In order to quantify this argument, we model the matter in the early Universe by a gas with $n_{\rm b}$ 
bosonic and $n_{\rm f}$ fermionic massless degrees of freedom, all at  the same
temperature. {\it In equilibrium} its energy density, pressure, and entropy density
are given by the usual relations ($n_{\rm eff}=n_{\rm b}+\frac{7}{8}n_{\rm f}$)
\begin{subequations}\label{reu:1.3}
\begin{align}
	\label{reu:1.3a}
	&\rho = 3\; p = (\pi^2/30) \, n_{\rm eff} \; T^4 \, ,\\[1mm] 
	\label{reu:1.3b}
	&s = (2\pi^2/45)\, n_{\rm eff} \; T^3 \,,\\[2mm]
	\intertext{so that in terms of $U\equiv \rho \; V$ and $S\equiv s\; V$,}
	\label{reu:1.3c}
	&T\; dS = dU + p\; dV \,.
\end{align}
\end{subequations}
In an out-of-equilibrium process of entropy generation the question arises 
how the various thermodynamical quantities are related then. To be as conservative as possible,
we make the assumption that the irreversible inflow of energy
destroys thermal equilibrium as little as possible in the sense that the equilibrium
relations (\ref{reu:1.3}) continue to be (approximately) valid. Such minimally non-adiabatic
processes were termed ``adiabatic'' (with the quotation marks) in ref.\ \cite{reu:lima1}.
\medskip

\noindent
\textbf{(5) Primordial entropy generation.}\index{entropy!primordial entropy generation}\index{primordial entropy generation}
Let us return to the modified continuity equation (\ref{reu:2.5b}). After multiplication by $a^3$
it reads
\be\label{reu:3.1}
[\dot\rho + 3H(\rho +p)] \; a^3 = \calp(t)\,,
\ee
where we defined
\be\label{reu:3.2}
\calp\equiv -\Big ( \frac{\dot{\lbar}+8 \pi\; \rho \; \dot{G}}{8\pi \; G} \Big ) a^3\,.
\ee
Without assuming any particular equation of state eq.\ (\ref{reu:3.1}) can be
rewritten as 
\be\label{reu:3.3}
\frac{d}{dt} (\rho a^3) +p\frac{d}{dt}(a^3) = \calp(t)\,.
\ee
The interpretation of this equation is as follows. Let us consider a unit {\it coordinate},
i.e., comoving volume\index{comoving volume} in the Robertson-Walker space-time. Its corresponding  
{\it proper} volume is $V=a^3$
and its energy contents is $U=\rho a^3$. The rate of change of these quantities is subject to 
(\ref{reu:3.3}): 
\be\label{reu:3.4}
\frac{dU}{dt}+p\frac{dV}{dt}=\calp(t)\,.
\ee
In classical cosmology where $\calp\equiv 0$ this equation together with the standard thermodynamic
relation $dU+pdV=TdS$ is used to conclude that the expansion of the Universe is adiabatic, i.e.,
the entropy inside a comoving volume does not change as the Universe expands, $dS/dt=0$.

When $\lbar$ and $G$ are time dependent, $\calp$ is nonzero and we interpret (\ref{reu:3.4})
as describing the process of energy (or ``heat'') exchange between the scalar fields $\lbar$ 
and $G$  and the ordinary matter. This interaction causes $S$ to change,
\be\label{reu:3.5}
T\frac{dS}{dt}=T\frac{d}{dt}(s a^3)=\calp(t)\,,
\ee
where here and in the following we write $S\equiv s \, a^3$ for the entropy carried by the matter inside
a unit comoving volume and $s$ for the corresponding proper entropy density.
The actual rate of change of the comoving entropy is 
\be\label{reu:3.6}
\frac{dS}{dt}=\frac{d}{dt}(s a^3)= \clp (t)\,,
\ee
where $\clp \equiv \calp /T$.
If $T$ is known as a function of $t$ we can integrate (\ref{reu:3.5}) to obtain $S=S(t)$. 
In the RG improved cosmologies the entropy production rate\index{entropy!entropy production rate} per comoving 
volume 
\be\label{reu:3.7}
\clp(t) = - \Big [ \frac{ \dot{\lbar}+8\pi \; \rho \; \dot G}{8\pi\;  G} \Big ] \frac{a^3}{T}
\ee
is nonzero because the gravitational ``constants''  $\lbar$ and $G$ have acquired a time 
dependence. 

Clearly we can convert the heat exchanged, $TdS$, to an entropy change only if the dependence
of the temperature $T$ on the other thermodynamical quantities, in particular $\rho$ and $p$
is known.  For this reason we shall now make the following assumption about the matter system and its
(non-equilibrium!)  dynamics:

{\it The matter system is assumed to consist 
of $n_{\rm eff}$ species of effectively massless degrees of freedom
which all have the same temperature $T$. The equation of state is $p=\rho/3$, 
\ie $w=1/3$, and $\rho$ depends on $T$ as 
\be\label{reu:3.9}
\rho(T) =\kappa^4 \; T^4 , \;\;\;\;\;\;\; \kappa\equiv (\pi^2 \; n_{\rm eff}/30)^{1/4}\,.
\ee
No assumption is made about the relation $s=s(T)$.}

The first assumption, radiation dominance and equal temperature, is plausible since we shall find that
there is no significant entropy production any more once $H(t)$ has dropped substantially below $\mp$.
The second assumption, eq.\ (\ref{reu:3.9}), amounts to the hypothesis formulated above, the approximate
validity of the {\it equilibrium} relations among $\rho$, $p$, and $T$.

Note that while we used (\ref{reu:1.3c}) in relating $\clp(t)$ to the entropy production and also
postulated eq.\ (\ref{reu:1.3a}), we do not assume the validity of the formula for
the entropy density, eq.\ (\ref{reu:1.3b}), a priori. We shall see that the latter is an automatic
consequence of the cosmological equations. To make the picture as clear as possible we shall
neglect in the following all ordinary dissipative processes in the cosmological fluid.

Using $p=\rho/3$ and (\ref{reu:3.9}) the entropy production rate can be seen to be a total time derivative,
$\clp(t) = {d}/{dt} \; [ \frac{4}{3} \; \kappa \; a^3 \; \rho^{3/4} ]$.
Therefore we can immediately integrate (\ref{reu:3.5}) and obtain 
\be\label{reu:3.22}
S(t)=\frac{4}{3}\; \kappa \; a^3\; \rho^{3/4} +S_{\rm c}\;,
\ee
or, in terms of the proper entropy density, 
$s(t) = \frac{4}{3} \; \kappa\; \rho(t)^{3/4} +\frac{S_{\rm c}}{a(t)^3}$.
Here $\ssc$ is a constant of integration. In terms of $T$, using 
(\ref{reu:3.9}) again, 
\be\label{reu:3.24}
s(t) = \frac{2\pi^2 }{45} \; n_{\rm eff} \; T(t)^3 +\frac{S_{\rm c}}{a(t)^3}\;.
\ee

The final result (\ref{reu:3.24}) is very remarkable for at least two reasons. First, 
for $\ssc=0$, eq.\ (\ref{reu:3.24}) has exactly the form (\ref{reu:1.3b}) which is valid for radiation
in equilibrium. Note that we did not postulate this relationship, only the $\rho(T)$--law
was assumed. The equilibrium formula $s\propto T^3$ was {\it derived} from the 
cosmological equations, i.e., the modified conservation law. This result makes the hypothesis
``non-adiabatic, but as little as possible'' selfconsistent.
Second, if $\lim_{t\rightarrow 0} \; a(t) \rho(t)^{1/4}=0$, which is  actually the
case for the most interesting class of cosmologies we shall find, then $S(t\rightarrow 0)=S_c$
by eq.\ (\ref{reu:3.22}). As we mentioned in the introduction, the most plausible initial value
of $S$ is $S=0$ which means a vanishing constant of integration $S_c$ here. But then,
with $S_c=0$, eq.\ (\ref{reu:3.22}) tells us that \emph{the entire entropy carried by the 
massless degrees of freedom today (CMBR photons) is due to the RG running}.
\medskip

\noindent
\textbf{(6) Entropy production for the RG trajectory realized by Nature.}\index{entropy production}
Substituting the NGFP solution \eqref{reu:NGFPsolution} for $w=1/3$ the entropy production rate \eqref{reu:3.7} reads
$\clp(t) = 4 \kappa \; (\alpha-1) \; A^3\; \widehat{\rho}^{3/4} \; t^{3\alpha -4}$.
For the entropy per unit comoving volume we
find, if $\alpha \not =1$, $S(t) = \ssc + \frac{4}{3} \kappa \; A^3 \; \widehat{\rho}^{3/4}\; t^{3(\alpha -1)}$,
and the corresponding proper entropy density is 
$s(t) =\ssc/(A^3 \; t^{3\alpha})+ 4\kappa \; \widehat\rho ^{3/4}/(3\;  t^3)$.
For the discussion of the entropy we must distinguish three qualitatively different cases.

\noindent
{\bf (i) The case $\mathbf{\alpha >1}$, i.e., $\mathbf {1/2<}\,\OLS \mathbf{<  1}$}: Here $\clp(t)>0$ so that the 
entropy and energy content of the matter system increases with time. By eq.(\ref{reu:3.7}), 
$\clp >0$ implies $\dot\lbar +8\pi  \rho  \dot G <0$. Since $\dot\lbar <0$ but $\dot G>0$
in the NGFP regime, the energy exchange is predominantly due to the decrease of $\lbar$
while the increase of $G$ is subdominant in this respect.
The comoving entropy $S(t)$ has a finite limit for $t\rightarrow 0$, 
$S(t\rightarrow 0) =\ssc$, and $S(t)$ grows monotonically for $t>0$. If $\ssc=0$,  
which would be the most
natural value in view of the discussion above, {\it all}  of the entropy carried
by the matter fields is due to the energy injection from $\lbar$. 

\noindent
{\bf (ii) The case $\mathbf{\alpha < 1}$, i.e., $\mathbf{ 0<}\,\OLS \mathbf{< 1/2}$}: 
Here $\clp(t)<0$ so that the energy and entropy of matter decreases. Since $\clp <0$ 
amounts to $\dot\lbar +8\pi  \rho  \dot G>0$, the dominant physical effect is the increase of 
$G$ with time, the counteracting decrease of $\lbar$ is less important.
The comoving entropy starts out from an infinitely positive value at the initial 
singularity, $S(t\rightarrow 0) \rightarrow +\infty$. This case is unphysical probably.

\noindent
{\bf (iii) The case $\mathbf{\alpha=1}$,  $\OLS\mathbf{=1/2}$}: Here $\clp(t)\equiv 0$, $S(t)=const$.
The effect of a decreasing
$\lbar$ and increasing $G$ cancel exactly.

At lower scales the RG trajectory leaves the NGFP and very rapidly ``crosses over'' to the GFP.
This is most clearly seen in the behavior of the anomalous dimension\index{anomalous dimension}
$\eta_{\rm N}(k)\equiv k\partial_k \ln G(k)$ which changes from its NGFP value $\eta_\ast =-2$ 
to the classical $\eta_{\rm N}=0$. This transition happens near $k\approx \mp$ or, since $k(t)\approx H(t)$,
near a cosmological ``transition" time $t_{\rm tr}$ defined by the condition 
$k(t_{\rm tr})=\xi H(t_{\rm tr})=\mp$. (Recall that $\xi=O(1)$.)
The complete solution to the improved equations can be found with numerical methods only. It 
proves convenient to use logarithmic variables normalized with respect to their respective values at the turning
point. Besides the ``RG time" 
$\tau \equiv \ln (k/\kat)$, we use $x \equiv \ln (a/\aT)$, 
$y \equiv \ln (t/\tT)$, and ${\cal U} \equiv \ln (H/\hT)$.

Summarizing the numerical results one can say  that for any value of $\OLS$ the UV cosmologies
consist of two scaling regimes and a relatively sharp crossover region near
$k,H\approx \mp$ corresponding to $x\approx -34.5$  
which connects them. At higher $k$-scales the fixed point approximation 
is valid, at lower scales one has a classical FRW cosmology in which $\lbar$
can be neglected. 

\begin{figure}[t]
\begin{center}
\includegraphics[width=.96\columnwidth]{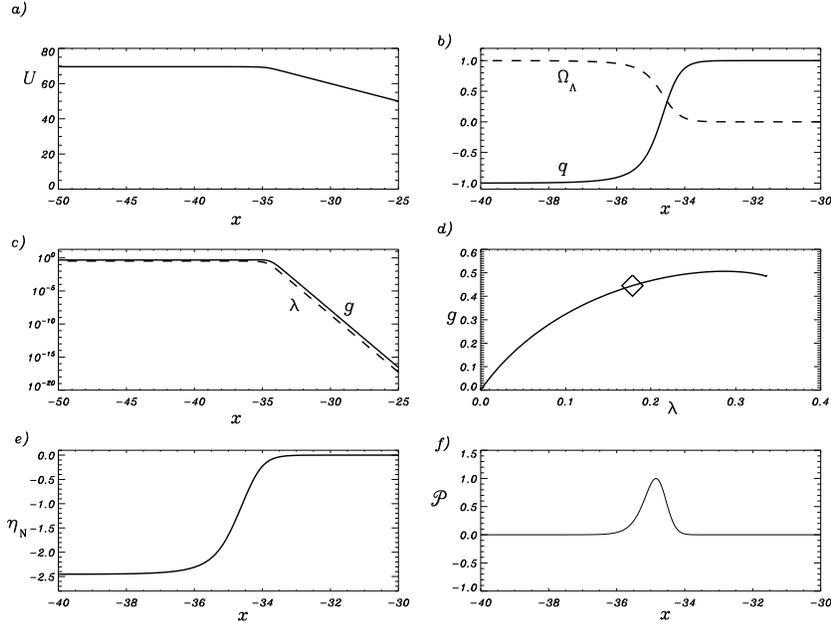}
\end{center}
\caption{The  crossover epoch of the cosmology for $\OLS=0.98$. The plots \textbf{a)}, \textbf{b)}, \textbf{c)} display the
	logarithmic Hubble parameter $\UU$, as well as $q$, $\OL$, $g$ and $\lambda$ as a function of the logarithmic
	scale factor $x$. A crossover is observed near $x\approx -34.5$. The diamond in plot \textbf{d)} indicates the point on
	the RG trajectory corresponding to this $x$-value. (The lower horizontal part of the trajectory is not visible on this
	scale.) The plots \textbf{e)} and \textbf{f)} show the $x$-dependence of the anomalous dimension and entropy production rate,
	respectively.}
\label{reu:fig7}
\end{figure}
As an example, Fig.~\ref{reu:fig7} shows the crossover cosmology with  $\OLS=0.98$ and $w=1/3$. 
The entropy production rate $\clp$ has a maximum at $t_{\rm tr}$ and quickly goes to zero for 
$t>t_{\rm tr}$; it is non-zero for all $t<t_{\rm tr}$. By varying the $\OLS$-value one can
check that the early cosmology is indeed described by the NGFP solution (\ref{reu:NGFPsolution}). For the logarithmic
$H$ vs. $a$- plot, for instance, it predicts $\UU=-2(1-\OLS)x $ for $ x<-34.4$. The left part of the plot in
Fig.~\ref{reu:fig7}a) and its counterparts with different values of $\OLS$ indeed comply with this relation.
If $\OLS \in (1/2,1)$ we have $\alpha = (2-2\OLS)^{-1}>1$ and $a(t)\propto t^\alpha$ describes a phase of
accelerated power law inflation. 

When $\OLS \nearrow 1$ the slope of $\UU (x) = -2 (1-\OLS)x$ decreases and
finally vanishes at $\OLS=1$. This limiting case corresponds  to a constant Hubble parameter, 
i.e., to de Sitter space. For values of $\OLS$ smaller than, but close to $1$ this 
de Sitter limit is approximated by an expansion $a\propto t^\alpha$ with a very large
exponent $\alpha$. The phase of power law inflation automatically comes to a halt 
once the RG running has reduced $\lbar$ to a value where the resulting vacuum energy density no longer
can overwhelm the matter energy density.

\section{Conclusions}
In these lectures we reviewed the basic ideas of Asymptotic Safety and explained why we believe that
Quantum Einstein Gravity is likely to be renormalizable in the modern non-perturbative sense.
We argued that the scale-dependence of the gravitational couplings intrinsic to Asymptotic Safety gives
rise to multifractal features of the effective space-times and should also have an impact on the
cosmological evolution of the Universe we live in. In the latter context, we proposed three possible candidate signatures:
 a period of automatic, cosmological constant-driven inflation that requires
no ad hoc inflaton, the entropy carried by the radiation
which fills the Universe today,
and the primordial density perturbations necessary for structure formation. If these
perturbations are an imprint of the metric fluctuations in the NGFP regime, the ``critical phenomenon''
properties of the latter might be the origin of the observed scale free spectrum of the former.
It is indeed an exciting idea that what we see when we look at the starry sky, during a clear summer
night on the Cycladic Islands, for instance, might actually be a snapshot of the geometry fluctuations
governed by the short distance limit of QEG, and tremendously magnified by the cosmic expansion.

\begin{acknowledgement}
	M.~R.\ would like to thank the organizers of the 6th Aegean Summer School for the opportunity to present
	this material and their cordial hospitality at Naxos. We are also grateful to D.~Benedetti and
	J.~Henson for sharing their Monte Carlo data with us, and A.~Nink for a careful reading of the manuscript.
	The research of F.S.\ is supported by the Deutsche Forschungsgemeinschaft (DFG)
    within the Emmy-Noether program (Grant SA/1975 1-1).
\end{acknowledgement}


%

\end{document}